\documentclass[useAMS,usenatbib]{mn2e}
\usepackage{caption}
\usepackage{subcaption}
\usepackage{multirow,multicol}
\usepackage{amsbsy,amsmath}
\usepackage{grffile}
\usepackage{graphics,graphicx}
\usepackage{subfig}
\usepackage{longtable}
\usepackage[latin1]{inputenc}
\usepackage{amsmath}
\usepackage{amsfonts}
\usepackage{amssymb}
\usepackage{makeidx}
\usepackage{float}
\restylefloat{figure}
\usepackage{stfloats}
\usepackage{epsfig}
\usepackage{graphicx}
\usepackage{caption}
\usepackage{color}
\usepackage{tabularx}
\usepackage{journal_names}
\usepackage{natbib}
\usepackage{multicol}
\usepackage{gensymb}
\usepackage[compact]{titlesec}
\usepackage{aas_macros}
\usepackage{comment}
\title[MACS clusters]{A study of diffuse radio sources and X-ray emission in six massive clusters}
\author[Parekh et al.]{Viral Parekh $^{1}$\thanks{E-mail:
viral@rri.res.in}, K.S. Dwarakanath$^{1}$, Ruta Kale$^{2}$, and Huib Intema$^{3}$ \\
$^1$Raman Research Institute, C. V. Raman Avenue, Sadashivnagar, Bangalore 560080 \\
$^2$National Centre for Radio Astrophysics, T. I. F. R., Post Bag 3, Ganeshkhind, Pune 411007\\
$^3$Leiden Observatory, Leiden University, PO Box 9513, 2300 RA Leiden, The Netherlands}

\begin{document}
\date{February 2016}
\label{firstpage}

\maketitle

\begin{abstract}

\par The goal of the present study is to extend our current knowledge of the diffuse radio source (halo and relic) populations to $z$ $>$ 0.3. Here we report GMRT and EVLA radio observations of six galaxy clusters taken from the MAssive Cluster Survey (MACS) catalogue to detect diffuse radio emission. We used archival GMRT (150, 235 and 610 MHz) and EVLA (L band) data and made images at multiple radio frequencies of the following six clusters - MACSJ0417.5-1154, MACSJ1131.8-1955, MACSJ0308.9+2645, MACSJ2243.3-0935, MACSJ2228.5+2036 and MACSJ0358.8-2955. We detect diffuse radio emission (halo or relic or both) in the first four clusters. In the last two clusters we do not detect any diffuse radio emission but we put stringent upper-limits on their radio powers. We also use archival {\it Chandra} X-ray data to carry out morphology and substructure analysis of these clusters. We find that based on X-ray data, these MACS clusters are non-relaxed and show substructures in their temperature distribution. The radio powers of the first four MACS clusters are consistent with their expected values in the $L_{x}$--$P_{1.4GHz}$ plot. {However, we found ultra-steep spectrum radio halo in the MACSJ0417.5-1154 cluster whose rest-frame cut-off frequency is at $\sim$ 900 MHz}. The remaining two clusters whose  radio powers are $\sim$ 11 times below the expected values are most likely to be in the `off-state' as has been postulated in some of the models of radio halo formation.

\end{abstract}


\section{Introduction}
\par Giant radio halos and relics are large ($\sim$ Mpc size) diffuse synchrotron sources observed in massive and merging galaxy clusters. These extended radio sources are associated with diffuse, non-thermal, and low surface brightness radio emission with steep spectra (-0.7 $ \gtrsim \alpha \gtrsim $ -1.4; S $\propto$ $\nu^{\alpha}$ where $\alpha$ is spectral index and S is the flux density at frequency $\nu$). Radio halos are generally centrally located in galaxy clusters and coincident with X-ray emission while radio relics are found at the peripheries of galaxy clusters (\citealt {2012A&ARv..20...54F}, and reference therein). 
\par The formation and evolution of diffuse radio sources are still not clear. Several theories have been made for the mechanism of transferring energy into the relativistic electron population to generate cluster-wide radio halos. The relativistic particles could be injected in the cluster volume by AGN activity such as quasars, radio galaxies, etc. This primary electrons population needs to be re-accelerated \citep{2001MNRAS.320..365B, 2001ApJ...557..560P} to compensate for radiative losses. A recent cluster merger is the most likely process to stimulate the re-acceleration of relativistic particles through injection of {turbulence} in the cluster volume \citep{2001MNRAS.320..365B,2001ApJ...557..560P,2008SSRv..134..207P}. Another model for radiating particles in radio halos incorporates secondary electrons as the result of inelastic nuclear collisions which have occurred between the relativistic protons and thermal ions of the ambient ICM. The protons diffuse on a large scale because their energy losses are negligible. They can continuously produce in-situ electrons which are distributed throughout the cluster volume \citep{1999APh....12..169B, 2001ApJ...562..233M}. It is possible that the high energy electrons, responsible for the synchrotron emission, arise from the decay of
secondary products of the neutralino annihilation in the dark matter halos of galaxy
clusters has also been proposed \citep{2001ApJ...562...24C}.
\par Different models have been suggested for the origin of the relativistic electrons radiating from relics. There is observational evidence that relics can be tracers of shock-waves in merger events \citep {1998A&A...332..395E, 2006Sci...314..791B}. These shocks, expanding with high velocity (Mach number $\sim$ 1-3), can accelerate electrons to high energies and compress magnetic fields, giving rise to synchrotron radiation being emitted from large regions \citep{2012A&ARv..20...54F}. The accelerated particles will have a power-law energy distribution, as well as magnetic fields aligned parallel to the shock front. This is in agreement with their elongated structure almost perpendicular to the merger axis \citep{2011MNRAS.418..230V,2011A&A...528A..38V}.
\par It has been shown that X-ray data are crucial in the study of non-thermal radio emission from galaxy clusters. A number of authors have shown the correlation between radio luminosity of a halo ($P_{1.4GHz}$) vs the host cluster X-ray luminosity ($L_{X}$), mass, and temperature \citep {2000ApJ...544..686L, 2001A&A...376..803G, 2003ASPC..301..143F, 2006MNRAS.369.1577C, 2007MNRAS.378.1565C, 2009A&A...507.1257G,2009A&A...507..661B, 2011MmSAI..82..499V,2011JApA...32..519C}. These correlations suggest that the most powerful {radio halos} are found in the biggest and hottest clusters with the greatest X-ray luminosity. Furthermore, these correlations show a close link between non-thermal and  thermal galaxy cluster physics. \cite {2001A&A...376..803G} showed the point-to-point correlation between X-ray and radio surface brightness for six {radio halo} clusters. This suggests a correlation between { radio halo} and X-ray emission in galaxy clusters. 
\par  {High resolution} X-ray data showed a relationship between non-thermal radio sources and X-ray cluster morphology \citep[][hereafter, P15]{2001ApJ...553L..15B, 2001A&A...378..408S, 2010ApJ...721L..82C, 2015A&A...575A.127P}. This indicates a connection between the cluster dynamical state and the occurrence of a radio halo. The X-ray surface brightness distribution of merging clusters typically exhibit non-concentric iso-intensity contours, small-scale substructures, multiple peaks and/or poor optical and X-ray alignments. \cite{2001ApJ...553L..15B} first showed a correlation between the power-ratio ($P_{1}/P_{0}$) calculated for {\it ROSAT} observed ($\sim$ 14) X-ray clusters and $P_{1.4GHz}$ measured for radio data. The power-ratios measure both the multi-pole expansion of a 2-D structure and dynamical state of a cluster \citep{1995ApJ...452..522B, 1996ApJ...458...27B}. Buote concluded that approximately $P_{1.4GHz}$ $\propto$ $P_{1}$/$P_{0}$, which means the clusters that host the most powerful {radio halos} are undergoing the greatest exodus from a virialized state. Recently, \cite {2010ApJ...721L..82C} used three parameters, namely centroid shift \citep {1993ApJ...413..492M, 2006MNRAS.373..881P, 2008ApJS..174..117M, 2010A&A...514A..32B}, third order power ratio ($P_{3}$/$P_{0}$) \citep{2012arXiv1210.6445W} and Concentration \citep {2008A&A...483...35S} to demonstrate a relationship between cluster mergers and the presence of a radio halo. The analysis of \cite {2010ApJ...721L..82C} supports the relationship between cluster dynamical activity (or dynamical state) and cluster-wide diffuse sources. This relationship, in turn, supports the particle (re-)acceleration mechanism in the formation of radio halos in galaxy clusters. The current scenario indicates that diffuse radio features (radio halos) are seen only in merging clusters. 
\subsection{MAssive Cluster Survey}
\par The Massive Cluster Survey (MACS) cluster sample comprises a total of 124 clusters in the redshift range 0.3 $<$ $z$ $<$ 0.7 \citep{2001ApJ...553..668E}. These clusters are spectroscopically confirmed and represent a statistically complete sample of X-ray luminous ($\sim$ 10$^{45}$ erg s$^{-1}$ (0.1--2.4 keV)), massive (10$^{14}$ $\sim$ 10$^{15}$ $M_{\odot}$), and distant clusters of galaxies. Many MACS clusters have been studied for diffuse radio sources for example MACSJ07175+3745, MACSJ1149.5+2223, MACSJ1752.1+4440, and MACSJ0553.4-3342 \citep{2009A&A...503..707B,2009A&A...505..991V,2012MNRAS.426...40B}. The detection probability of radio halo is rather small ($<$ 10$\%$) for all clusters, the probability increases to $\sim$ 40$\%$ for clusters with $L_{x}$ $>$ 10$^{44}$ erg s$^{-1}$ \citep{2008A&A...484..327V}. Under these circumstances, majority of MACS clusters might be expected to host detectable radio halos or relics or both.  With these criteria of most massive and luminous clusters, we selected a sample of six most massive ($>$ 6 $\times$ 10$^{14}$ $M_{\odot}$) and disturbed galaxy clusters from the MACS catalogue \citep{2010MNRAS.407...83E}. The Sunyaev-Zel'dovich (SZ) mass of these clusters are also very high which suggests that they are good candidates to host diffuse radio sources \citep{2013ApJ...777..141C}. In this work, we studied these clusters, in radio and X-ray observations, to detect diffuse radio sources and its connection with cluster merger. We list the sample of MACS clusters in Table \ref{MAC_clus_prop} with other X-ray properties. 
\par This paper is organized as follows. \S\ 2 gives a brief overview of radio observations and data reduction. \S\ 3  gives the details about individual clusters in the sample. In \S\ 4, we present X-ray data, temperature map, and morphology analysis. \S\ 5 gives substructure analysis of six radio halo clusters. Finally, \S\ 6 gives discussion and conclusions. We assumed $H_{0}$ = 70 km s$^{-1}$ Mpc$^{-1}$ $\Omega_{M}$ = 0.3 and $\Omega_{\Lambda}$ = 0.7 throughout the paper.
 
\begin{table*}
 \centering
 \caption{MACS clusters and its properties. Column (1) cluster name, (2) alternate name, (3) right ascension, (4) declination, (5) redshift, (6) angular to linear scale, (7) SZ mass of the cluster. We used the latest 2015 Planck catalogue available at the Planck Legacy Archive at http://http://pla.esac.esa.int/pla/catalogues, (8) temperature taken from \protect \cite{2008ApJ...682..821C} and (9) luminosity (0.1-2.4 keV) calculated based on $f^{x}_{r500,CXO}$ given in \protect \cite{2010MNRAS.407...83E}.}
 \label{MAC_clus_prop}
 \begin{tabular}{ccccccccccc}
  \hline

  MACS clusters & alt. name &R.A.  & Declination & $z$& scale &$Y{_{SZ,M_{500}}}$ &T & $L_{x}$ \\
             & &$J2000$ & $J2000$ &  &kpc/$''$ &10$^{14}$ $M_{\odot}$ &keV & 10$^{45}$ (erg s$^{-1}$)\\
  \hline
  MACSJ0417.5-1154 &  &04 17 34.6 &-11 54 32  &0.44 &5.68&    12.25$^{+0.52}_{-0.55}$& 11.07 & 3.66 \\  
  MACSJ1131.8-1955 & A1300 &11 32 00.7 &-19 53 34  &0.30 &4.45  & 8.97 $^{+0.45}_{-0.45}$& 7.75 & 1.54\\ 
  MACSJ0308.9+2645 &  &03 08 55.9 &+26 45 38  &0.35 &4.94  & 10.75 $^{+0.62}_{-0.65}$& 10.54 & 1.79\\ 
  MACSJ2243.3-0935 &  &22 43 21.1 &-09 35 43  &0.44 &5.68  & 10.0 $^{+0.43}_{-0.44}$& 7.98 & 1.90\\   
  MACSJ2228.5+2036 &RXJ2228.6+2037  &22 28 37.1 &+20 36 31  &0.41&5.45 & 8.26 $^{+0.43}_{-0.44}$& 8.16 & 1.63\\ 
  MACSJ0358.8-2955 & A3192 &03 58 42.5 &-29 54 30  &0.42 &5.53 & 7.20 $^{+0.52}_{-0.49}$& 8.8 & 2.35\\  
  \hline
 \end{tabular}
\end{table*}

\section{Radio observations and Data reduction }
\subsection{610 and 235 MHz GMRT data}
\par We used GMRT archival data to image six MACS clusters. These clusters were observed during October, 2010. Each of the MACS cluster was observed for a total duration of $\sim$ 8 hours in the dual frequency (235/610 MHz) mode. This allows to record one polarization at each frequency band. The GMRT software back-end with a bandwidth of 6 and 32 MHz at 235 and 610 MHz frequencies, respectively were used in this observations. Data were analysed using the Astronomical Image Processing System (AIPS) and Common Astronomy Software Applications (CASA) packages (developed by NRAO). The data were inspected for the RFI (radio frequency interference), non-working antennas, bad baselines, channels and time period. Corrupted data were excised from the $u$-$v$ dataset. The flux density of each primary or flux calibrator(s) is set according to the \cite{1977A&A....61...99B}. The same calibrator was used for the bandpass calibration followed by determining the flux density of the secondary or phase calibrator(s) using the antenna gain solutions. Calibrated visibilities then used to create the images using the standard Fourier transform deconvolution method. Few rounds of self-calibration (3 phase + 1 amplitude) were applied to reduce the effects of residual phase errors in the data and to improve the quality of the final images. Wide field imaging with 256 wprojection planes were used in the CASA task `clean' (see CASA manual). Images were produced with a variety of weighting (uniform and natural) schemes for the visibilities.
   
\subsection{150 MHz TGSS data}
The TGSS\footnote{TIFR GMRT Sky Survey; see http://tgss.ncra.tifr.res.in} is a fully observed yet largely unreleased survey of the radio sky at 150 MHz as visible from the GMRT, covering the full declination range of -55$\deg$ to +90$\deg$. All these observational Data was recorded in half polarization (RR, LL) every 2~seconds in 256 frequency channels across 16 MHz of bandwidth (140--156~MHz). Each pointing was observed for about 15 minutes, split over 3 or more scans spaced out in time to improve UV-coverage. As a service to the community, this archival data has been processed with a fully automated pipeline based on the SPAM package \citep{2009A&A...501.1185I,2014arXiv1402.4889I}, which includes direction-dependent calibration, modeling and imaging to suppress mainly ionospheric phase errors. In summary, the pipeline consists of two parts: a \emph{pre-processing} part that converts the raw data from individual observing sessions into pre-calibrated visibility data sets for all observed pointings, and a \emph{main pipeline} part that converts pre-calibrated visibility data per pointing into stokes I continuum images. The flux scale is set by calibration on 3C48, 3C147 and 3C286 using the models from \citet{2012MNRAS.423L..30S}. More details on the processing pipeline and characteristics of the data products will be given in an upcoming paper on the first TGSS alternative data release \citep[ADR1]{2016arXiv160304368I}. For this study, ADR1 images were used to create mosaics at the cluster positions. All 150 MHz images, in out study, have a resolution of 20$''$ $\times$ 20$''$. 
\subsection{L band EVLA data}
\par We used EVLA archival observations for the sample of MACS clusters. In the archive, we found L band (1400 MHz), C configuration (max baseline 3.4 km) observations 11B-018 (February, 2012) for only two clusters: MACSJ0417.5-1154 and MACSJ2243.3-0935. These clusters were each observed for a duration of $\sim$ 1.5 hrs. These observations were carried out with 16 spectral windows (spws) from 1194 to 1892 MHz. Each spectral window has 64 channels. About half of the data were corrupted so we removed it from the $u$-$v$ data. We followed the standard procedure of data analysis. The flux density of each primary or flux calibrator(s) is set according to the \cite{2013ApJS..204...19P}. The calibrated visibilities were then imaged in CASA in the same way as described for 235 and 610 MHz observations. In further analysis, we convolved all images (235, 610 and 1575 MHz) to 20$''$ $\times$ 20$''$ resolution, which is the resolution of the 150 MHz TGSS images.
\par In order to make an error estimate in the flux density measurements, here we outline the procedure. There are two primary sources of errors in the flux density measurements: (1) an error due to the uncertainties in the flux densities of the unresolved source(s) used for calibration of the data. We assumed this error to be $\sim$ 10$\%$. and (2) since diffuse radio sources are extended sources, the errors in their flux density estimations will be the rms in the image multiplied by the square root of the ratio of the solid angle of the source to that of the synthesised beam. These two sources of errors are unrelated, so they are added in quadrature to estimate the final error on the flux densities of the extended sources, as shown below.

\begin{equation}
\Delta{S} = [(\sigma_{amp}S)^2 + (\sigma_{rms}\sqrt{n_{beams}})^2)]^{1/2},
\label{cal_err}
\end{equation}
  where $S$ is a flux density, $\sigma_{rms}$ is the image rms noise, and $n_{beams}$ is the number of beams in the extent of the source.

\section{Individual cluster}
\subsection{MACSJ0417.5-1154}
\par
\begin{figure*}[t]
    \centering
    \begin{subfigure}[t]{0.38\textwidth}
        \includegraphics[width=\textwidth]{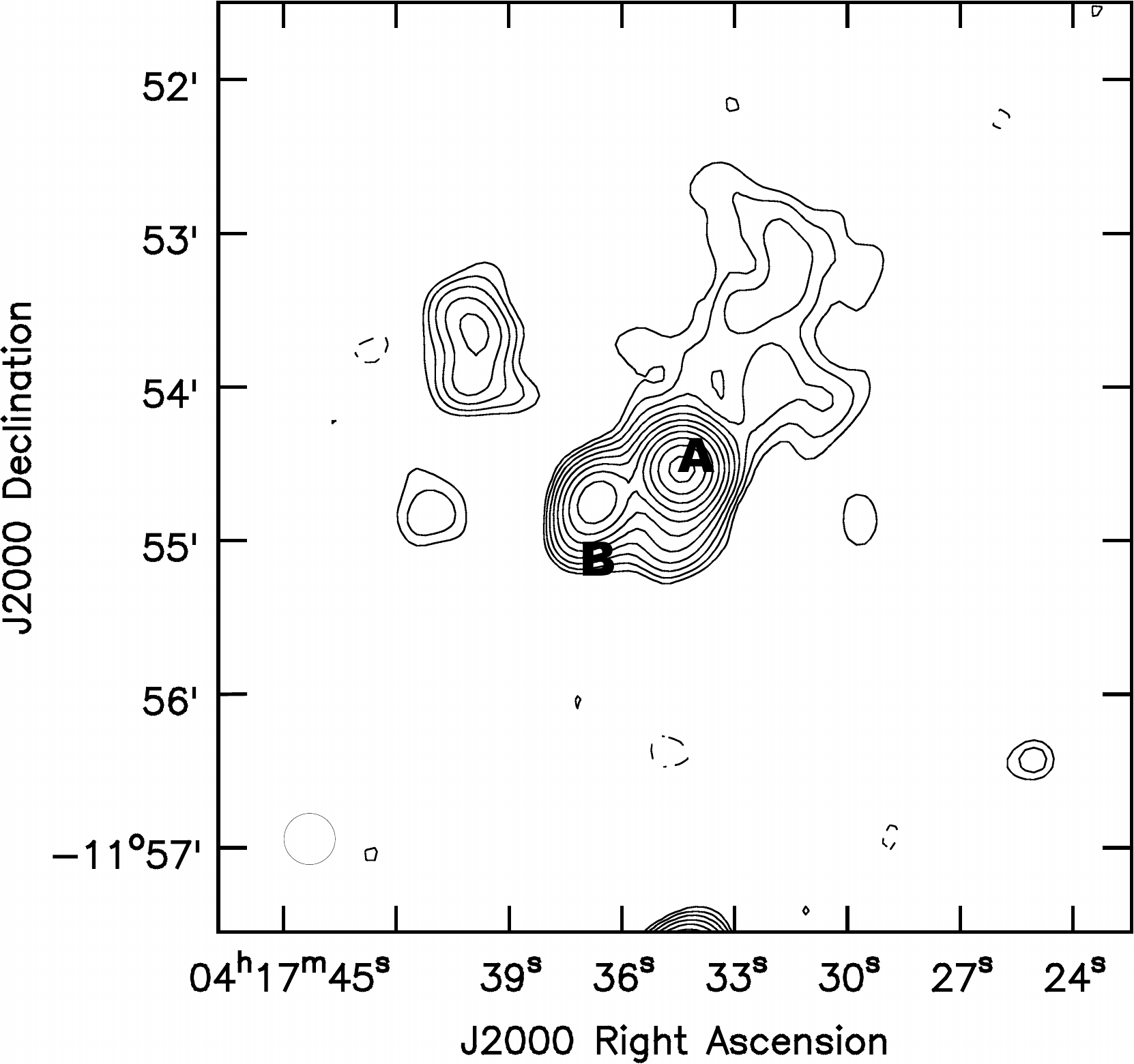}
        \caption{}
        \label{rfidtest_xaxis1}
    \end{subfigure}
    \begin{subfigure}[t]{0.39\textwidth}
        \includegraphics[width=\textwidth]{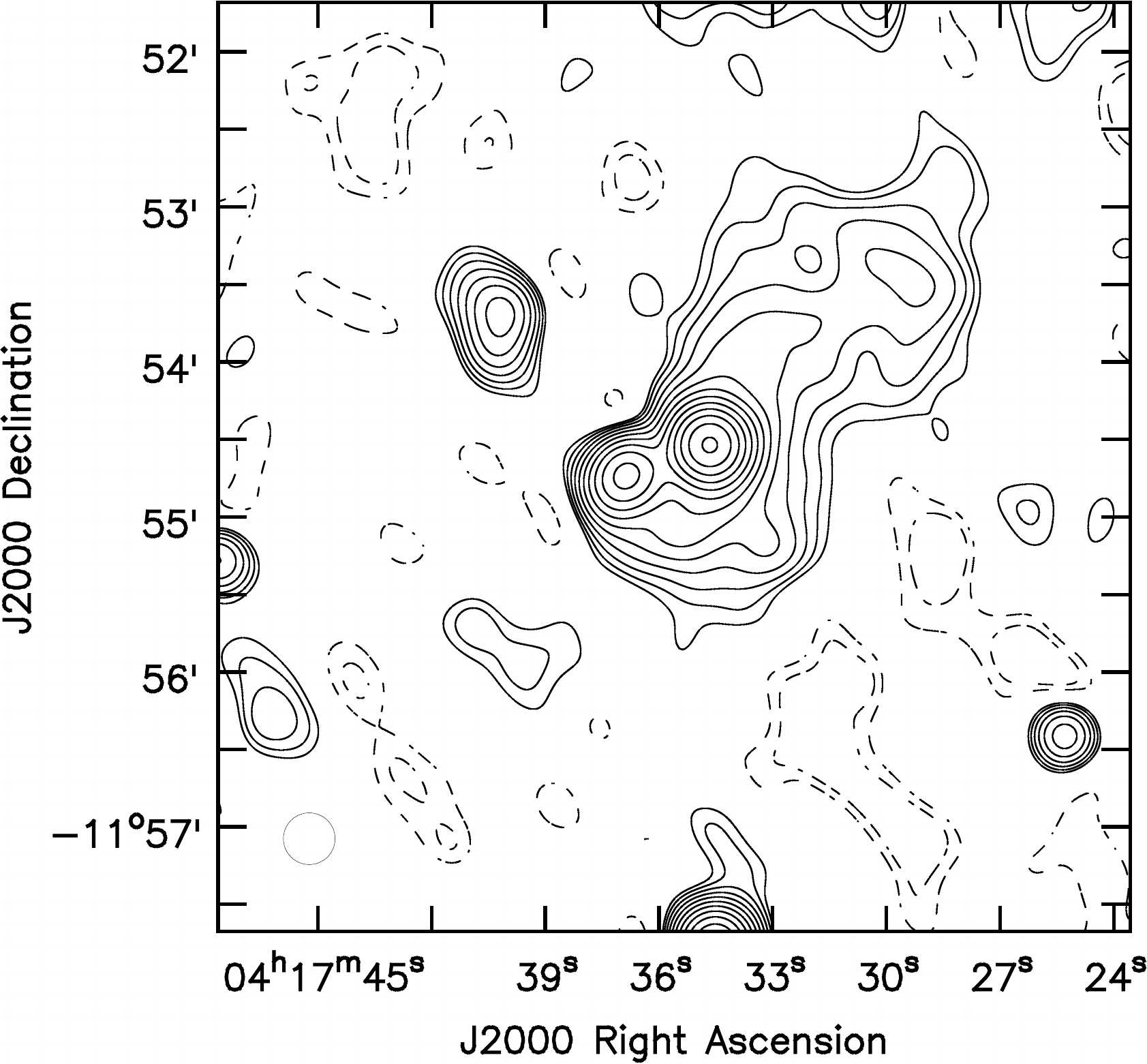}
        \caption{}
        \label{rfidtest_yaxis2}
    \end{subfigure}
    \begin{subfigure}[t]{0.38\textwidth}
        \includegraphics[width=\textwidth]{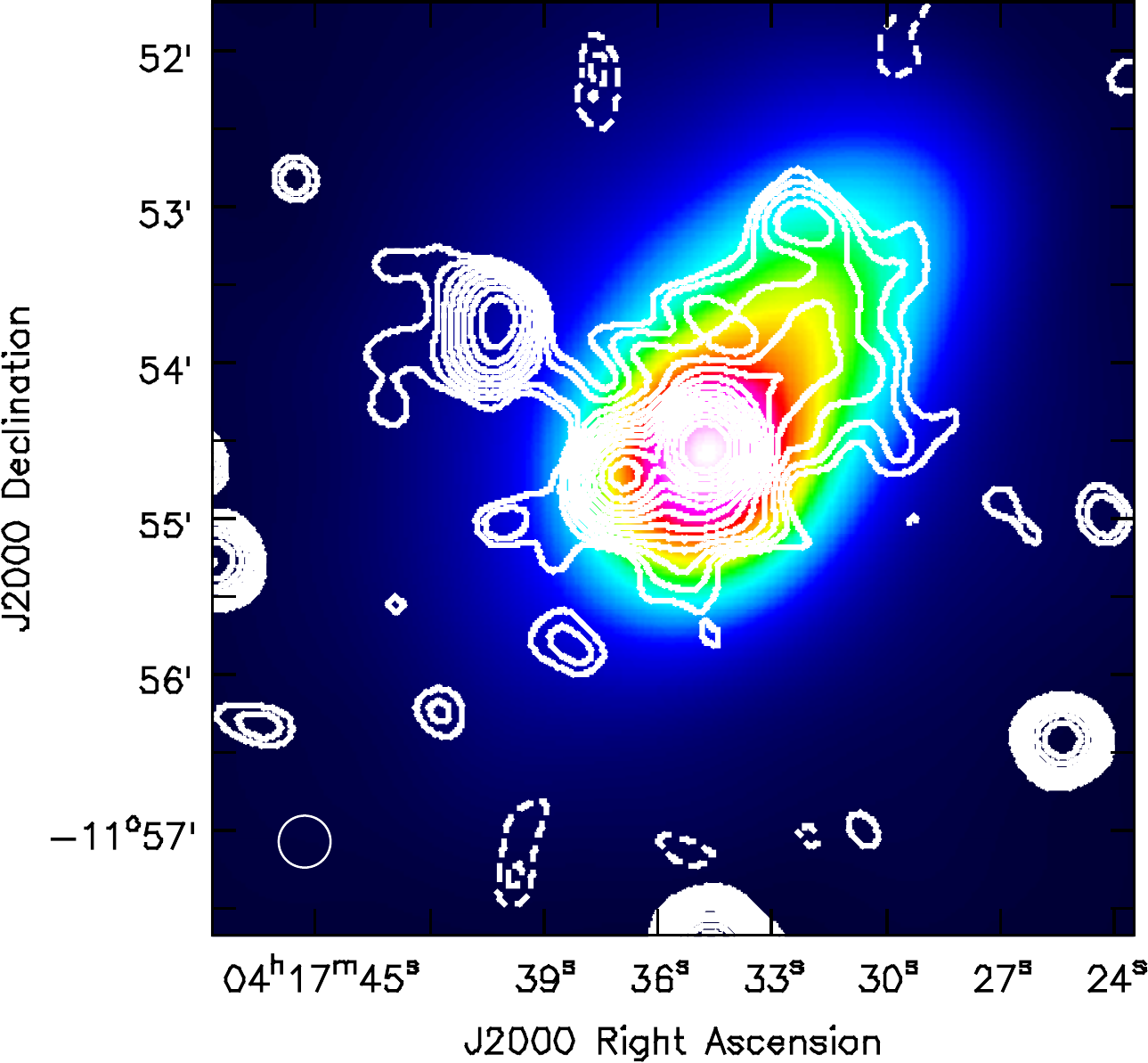}
        \caption{}
        \label{rfidtest_zaxis3}
    \end{subfigure}
        \begin{subfigure}[t]{0.45\textwidth}
        \includegraphics[width=\textwidth]{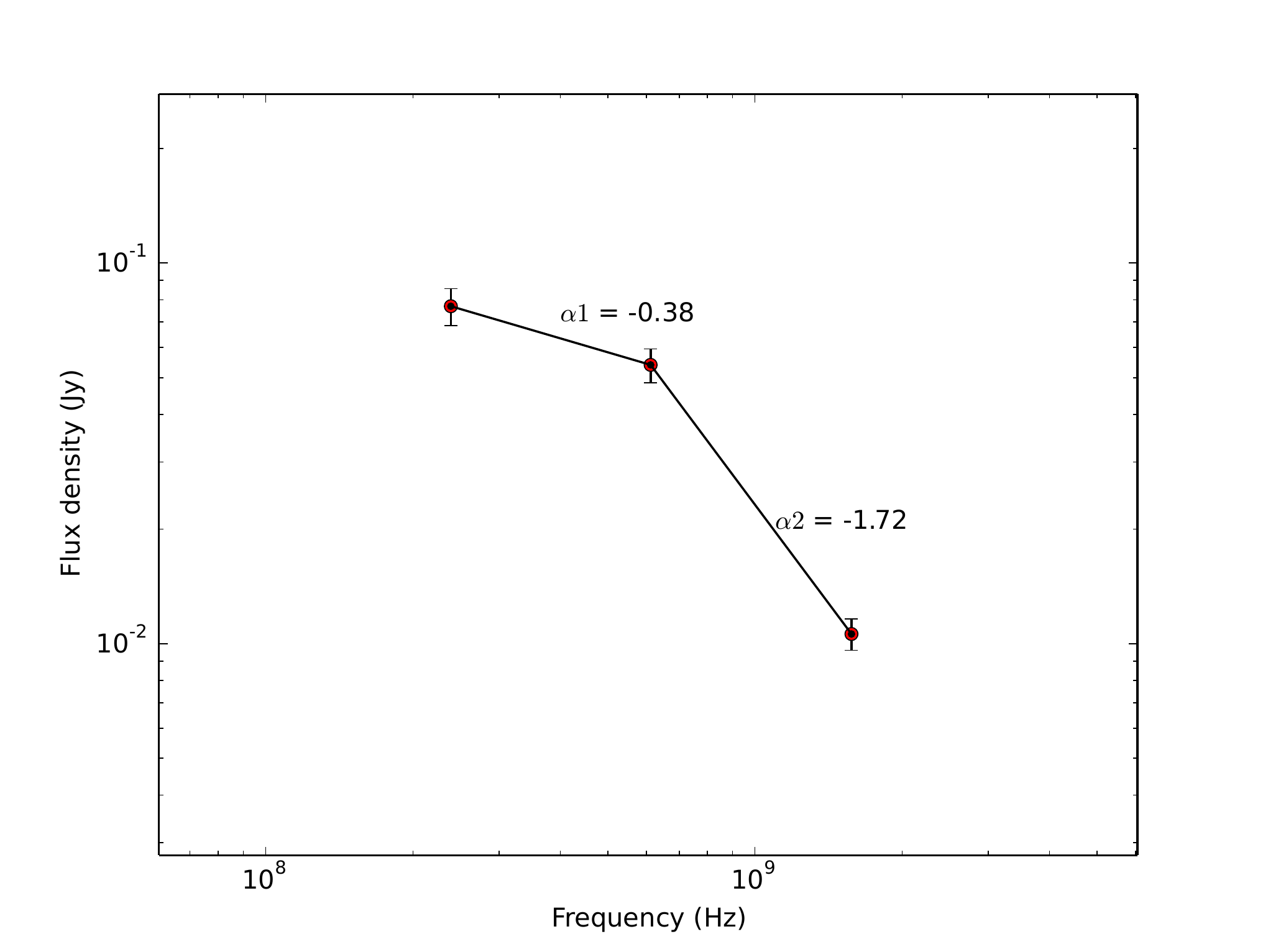}
        \caption{}
        \label{MAC0417_halo_spectra}
    \end{subfigure}
    \caption[]{ MACSJ0417-1155. (a) GMRT 235 MHz contours, (b) GMRT 610 MHz contours, (c) EVLA 1575 MHz contours on {\it Chandra} X-ray image, and (d) the integrated spectrum of the radio halo in MACSJ0417-1155. All radio images have a resolution of 20$''$ $\times$ 20$''$.  First contour is drawn at 3$\sigma$ where $\sigma_{235 MHz}$ = 0.44 mJy beam$^{-1}$, $\sigma_{610 MHz}$ = 0.15 mJy beam$^{-1}$, and  $\sigma_{1575 MHz}$ = 40 $\mu$Jy beam$^{-1}$. Contour level increases in steps of $\sqrt{2}$. Dashed line shows -3$\sigma$ contours. A colour version of this figure is available online.}
    \label{MAC0417_radio_img}
\end{figure*}

 MACSJ0417.5-1154 is a hot ($\sim$ 11 keV), X-ray luminous (3.66 $\times$ 10$^{45}$ erg s$^{-1}$) and massive cluster (12.25 $\times$ 10$^{14}$ $M_{\odot}$) at a redshift of $z$ = 0.44. MAC0417.5-1155 is the most massive cluster in the sample. The detection of a halo in this cluster was first reported by \cite{2011JApA...32..529D}. Figure~\ref{MAC0417_radio_img} shows the (a) 235 MHz, (b) 610 MHz GMRT radio images, and (c) 1575 MHz EVLA radio contours overlaid on the {\it Chandra} X-ray image of MACSJ0417.5-1155. In all these frequencies, there is extended, comet-like diffuse radio emission visible along the north-west direction. As shown in Figure~\ref{MAC0417_radio_img} (c), diffuse radio emission is coincident with the corresponding hot X-ray distribution in direction and size. Elliptical X-ray morphology clearly suggests that it is a disturbed and merging cluster. In 610 MHz observation, the largest linear size of the halo is $\sim$ 1 Mpc which is the typical size of radio halos in massive clusters. In the X-ray image, bright and compact core is visible which is situated with the optical core of one of the two merging clusters \citep[][hereafter, AM12]{2012MNRAS.420.2120M}. Further, this morphology is also similar to those observed in merging clusters such as A2146 and Bullet clusters, which indicates a high velocity merger [AM12]. 

\par Table \ref{MAC_clusters_rad_obs} contains flux densities and linear sizes of halos (within 3$\sigma$ contours) measured at 235, 610, and 1575 MHz. There are two unresolved point sources (A and B) {embedded in the diffuse radio emission} (Figure~\ref{MAC0417_radio_img}). To calculate the integrated flux density of the radio halo, we subtracted flux densities of these point sources (Table \ref{unresoved_src}) {which are calculated by Gaussian fitting}, from the total halo flux density measured within the given size at each frequency in Table \ref{MAC_clusters_rad_obs}. Figure~\ref{MAC0417_radio_img} (d) shows the spectrum\footnote{In this work we scaled all values of flux densities of halos and relic to the Baars scale according to the \protect \cite{1977A&A....61...99B}.}of the radio halo of MACSJ0417.5-1155.  {Between 235 and 610 MHz, the spectrum is flat ($\alpha$1 = -0.38) and between 610 and 1575 MHz the spectrum is steep ($\alpha$2 = -1.72).}  We also compared EVLA 1575 MHz flux density measurements with the L-band NVSS \citep {1998AJ....115.1693C} observation of MACSJ0417.5-1154. In the NVSS image, A and B point sources blended, so we calculated flux density for the whole halo region and then subtracted flux densities of A and B derived from the EVLA observation with assuming spectral index of 0.7. The total flux density of the halo is $\sim$ 18 mJy which is comparable with the EVLA result. {Spectral index images of radio halos are very important to understand the presence of any magnetic field gradients and/or spatial variations of the relativistic electrons high-energy cut-off. But, signal to noise ratios in these radio images are not adequate to produce spectral index map. }
There is no diffuse emission visible in the 150 MHz TGSS data. The expected surface brightness ($\sim$ 2 mJy beam$^{-1}$) is below the rms ($\sim$ 4 mJy beam$^{-1}$) of 150 MHz observation. 
\par {We estimate the equipartition magnetic field \citep[see][for reviews]{2002ARA&A..40..319C,2004IJMPD..13.1549G,2008SSRv..134...93F} in the MACSJ0417-1155 radio halo to be $\sim$ 1.0 $\mu$G.}
 
\subsection{MACSJ1131.8-1955}
\par MACSJ1131.8-1955 or A1300 ($z$ = 0.30) is a complex and post-merging cluster as evident by both X-ray and optical observations \citep{1997A&A...326...34L, 2012MNRAS.420.2480Z}. Its X-ray luminosity is 1.54 $\times$ 10$^{45}$ erg s$^{-1}$ and mass is 7.75 $\times$ 10$^{14}$ $M_{\odot}$. First detection of the radio halo and relic in MACSJ1131.8-1955 was reported by \cite{1999MNRAS.302..571R}. They have detected diffuse radio sources at higher frequencies (1.34, 2.4, 4.8 and 8.6 GHz) using ATCA and MOST radio telescopes. It was also detected in the 325 MHz GMRT observations \citep{2013A&A...551A..24V}. Here we report halo and relic detection at 150, 235 and 610 MHz observations using the GMRT.
\par
\begin{figure*}[t]
    \centering
    \begin{subfigure}[t]{0.38\textwidth}
        \includegraphics[width=\textwidth]{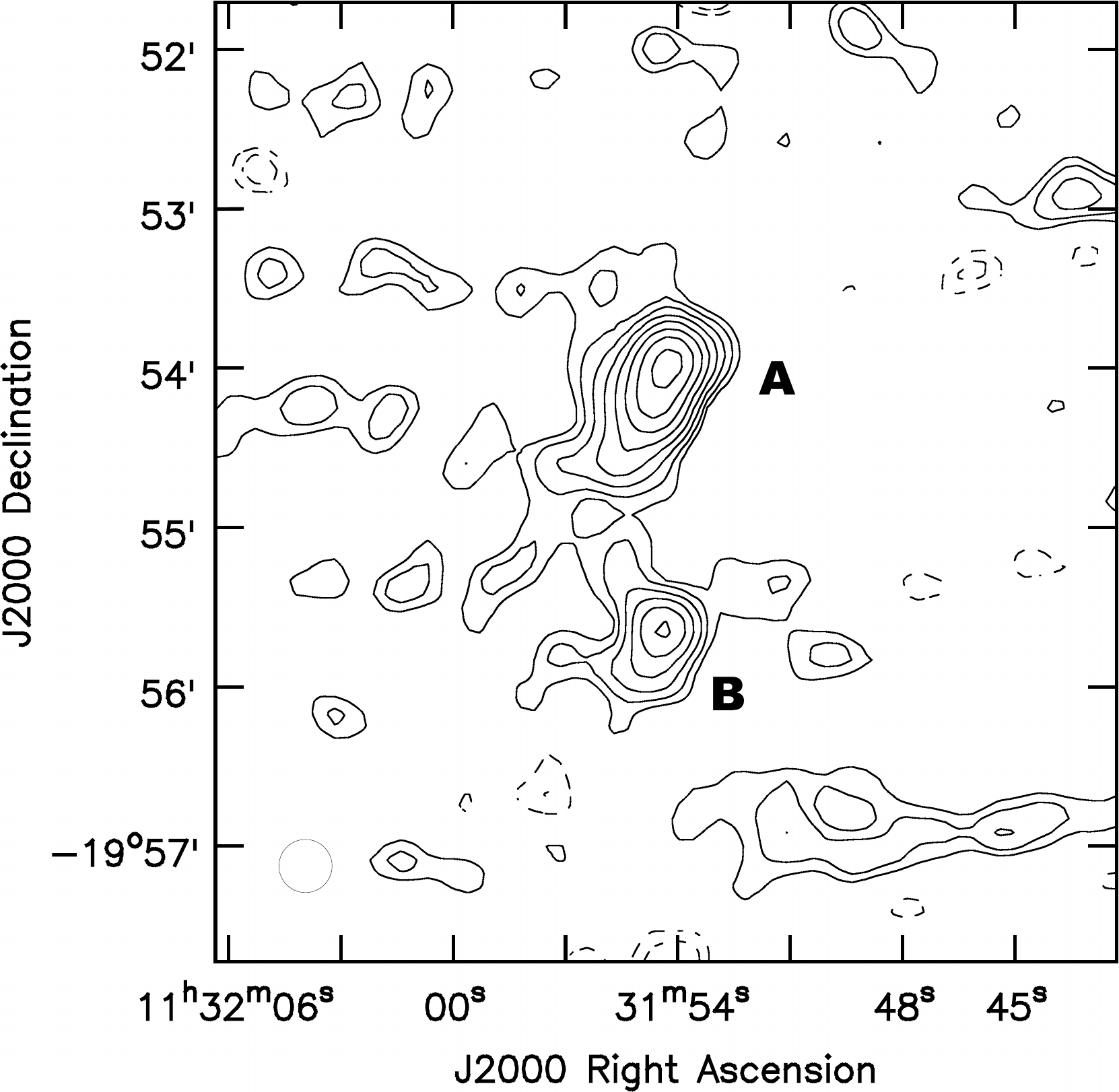}
        \caption{}
        \label{rfidtest_xaxis}
    \end{subfigure}
    \begin{subfigure}[t]{0.38\textwidth}
        \includegraphics[width=\textwidth]{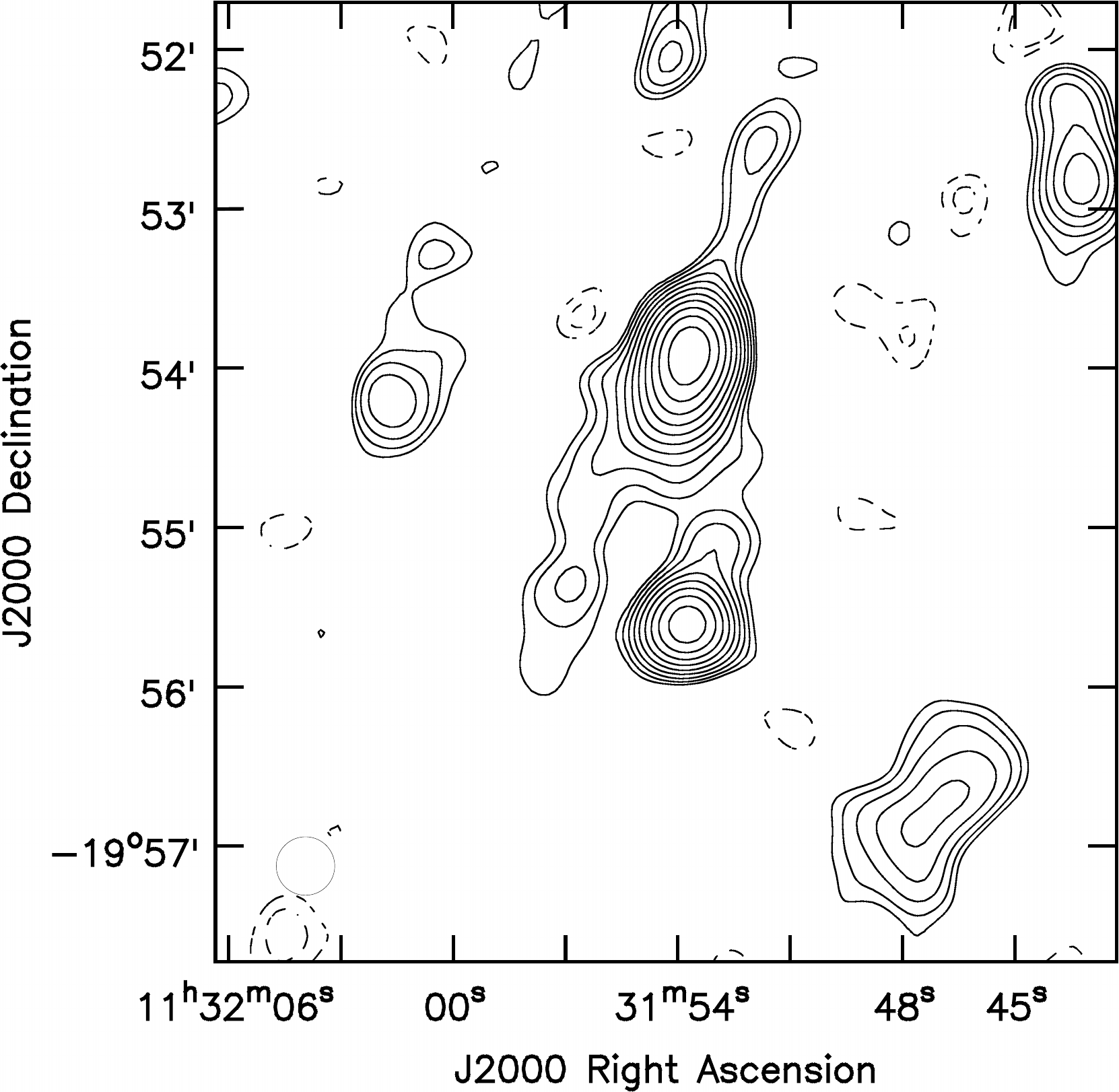}
        \caption{}
        \label{rfidtest_yaxis}
    \end{subfigure}
    \begin{subfigure}[t]{0.38\textwidth}
        \includegraphics[width=\textwidth]{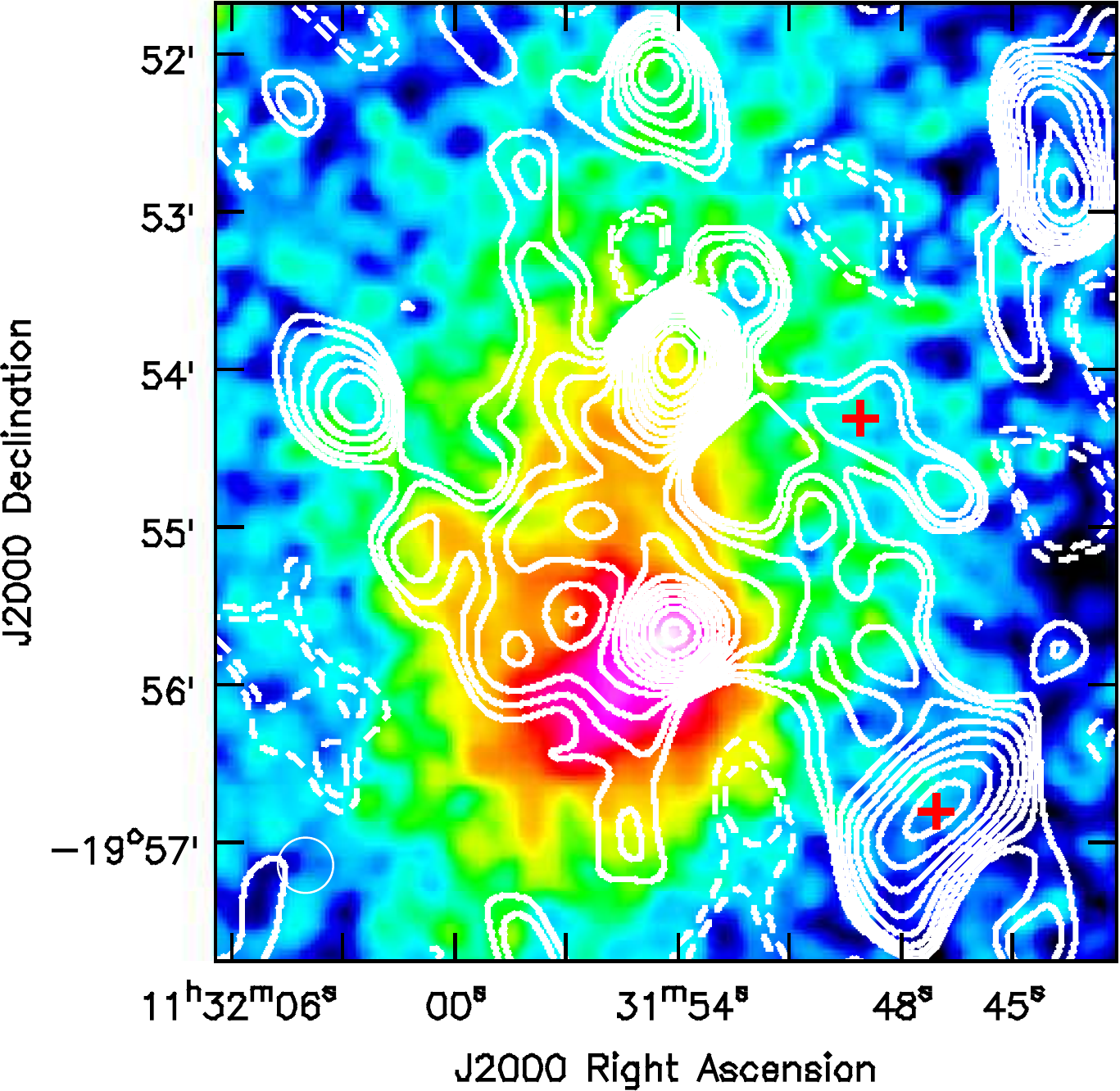}
        \caption{}
        \label{rfidtest_zaxisbridge}
    \end{subfigure}
    \caption[]{MACSJ1131.8-1955. (a) TGSS 150 MHz contours, (b) GMRT 235 MHz contours, and (c) GMRT 610 MHz contours on {\it Chandra} X-ray image. All radio images have a resolution of 20$''$ $\times$ 20$''$. First contour is drawn at 3$\sigma$ where $\sigma_{150 MHz}$ = 5 mJy beam$^{-1}$, $\sigma_{235 MHz}$ = 3 mJy beam$^{-1}$, and  $\sigma_{610 MHz}$ = 0.30 mJy beam$^{-1}$. Contour level increases in steps of $\sqrt{2}$. {\color{red} +}  sign are for the relics sources. Dashed line shows -3$\sigma$ contours.}
    \label{MACSJ1131_radio_img}
\end{figure*}

  Figure~\ref{MACSJ1131_radio_img} shows the (a) 150 MHz TGSS observations, (b) 235 MHz GMRT and, (c) 610 MHz GMRT observation superimposed on the {\it Chandra} X-ray image of MACSJ1131.8-1955.  In all three frequencies radio halo and south-west relic are detected. This south-west relic is marked with `{\color{red}+}' in Figure~\ref{MACSJ1131_radio_img} (c). As we noticed, in all three bands, radio halo extends mainly in the east direction of the peak in the X-ray surface brightness distribution. Furthermore, radio halo looks more extended (in the east direction) in 610 MHz data as compared to the 150 and 235 MHz observations. It has very disturbed and elongated X-ray morphology which indicates a series of current mergers. The overall shape and size of both halo and relic are comparable to the latest findings in 325 MHz observation. We detected largest linear sizes of halo and relic at 610 MHz are $\sim$ 815 kpc and $\sim$ 446 kpc, respectively which is comparable to the largest linear sizes of halo $\sim$ 890 kpc and relic $\sim$ 450 kpc at 325 MHz, respectively. In 150 MHz observation, we detected elongated structure (above 3$\sigma$ contour) at the position of south-west relic as seen in 235 and 610 MHz observations. 
We marked two unresolved point sources (A and B) in the 150 MHz image which are also visible at the same position in 235 and 610 MHz images. Since diffuse radio emission is very complex, it is difficult to avoid these point sources to calculate total flux density of the radio halo.  
The integrated flux densities (after subtracting flux densities of point sources) of the radio halo and relic, within the 3$\sigma$ contour level, are reported in Table \ref{MAC_clusters_rad_obs}. 

\begin{figure*}
\centering
\includegraphics[scale=0.40]{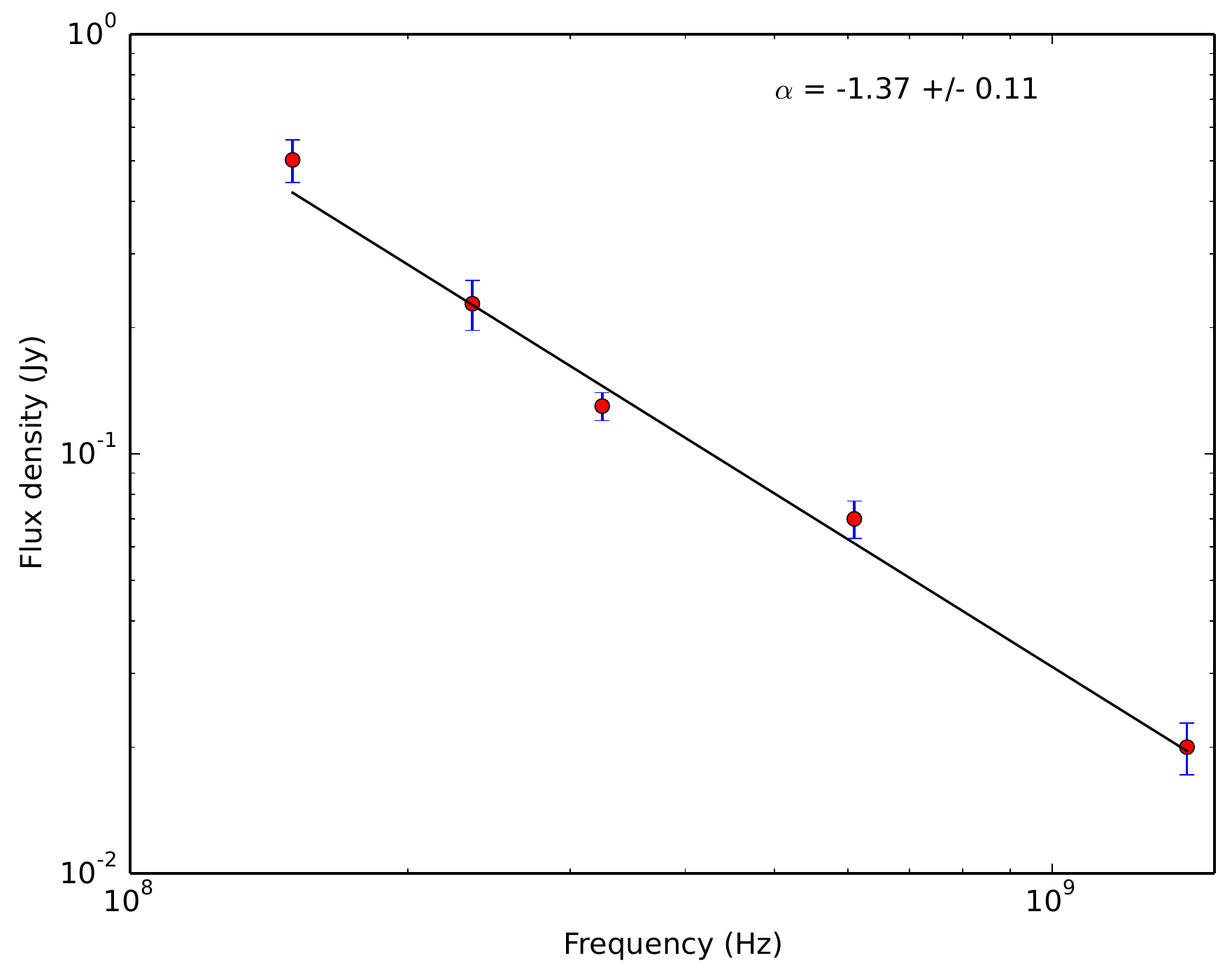}
\includegraphics[scale=0.40]{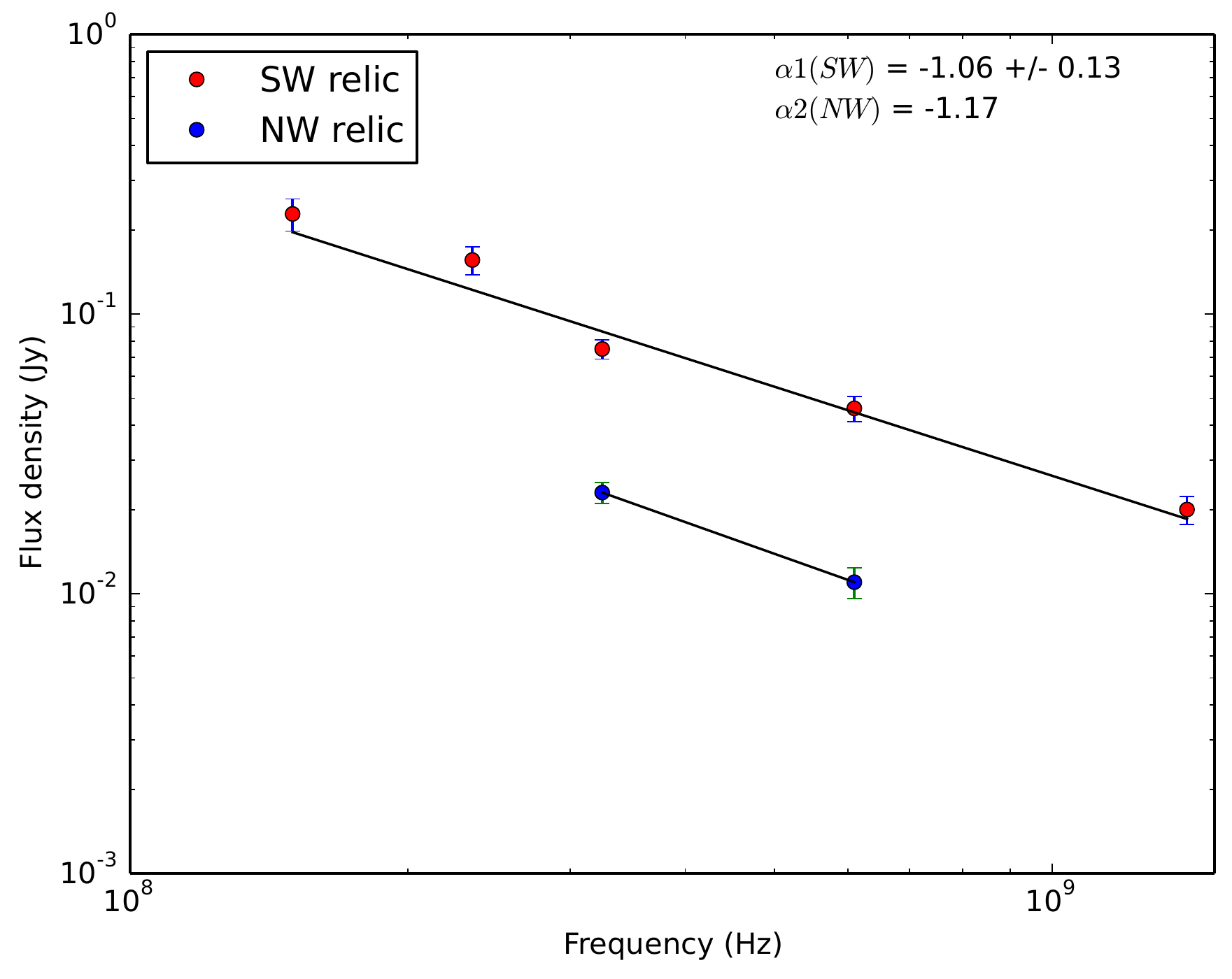}
\caption {The integrated spectrum of halo (left) and relic(s) (right) of MACSJ1131.8-1955.}
\label{MACSJ1131_halo_relic_spectra}
\end{figure*}

We show halo and relics spectra in Figure~\ref{MACSJ1131_halo_relic_spectra}. We used NVSS flux density (20 mJy) of the radio halo and relic. The integrated spectral index, for the halo is, $\alpha_{\small {150MHz}}^{\small{1400MHz}}$ = -1.37 $\pm$ 0.11 which is comparable with finding of \cite{2011MmSAI..82..541G}. If we use ATCA flux density (10 mJy) then $\alpha_{\small {150MHz}}^{\small{1400MHz}}$ = -1.72 $\pm$ 0.12. Spectral index for the south-west relic is, $\alpha_{\small {150MHz}}^{\small{1400MHz}}$ = -1.06 $\pm$ 0.13 which is comparable with the previous findings \citep{2011MmSAI..82..541G, 2013A&A...551A..24V}. If we use ATCA flux density (15 mJy) then $\alpha_{\small {150MHz}}^{\small{1400MHz}}$ = -1.19 $\pm$ 0.05.
\par \cite{2013A&A...551A..24V} have noticed candidate relic ($\sim$ 700 kpc size) in 325 MHz image, based on its overall morphology and location, at the north-west direction from the centre of cluster. We also noticed this candidate relic in 610 MHz observation which is also marked with `{\color{red}+}' sign. We measured the total flux density of this candidate relic to be 11 $\pm$ 1.4 mJy and the largest linear size is $\sim$ 670 kpc which are comparable with the previous results. We show its spectrum between 325 and 610 MHz in Figure~\ref{MACSJ1131_halo_relic_spectra}. The spectrum ($\alpha$ = -1.17) of north-west relic is steeper than south-west relic. This candidate relic is not visible in 150 and 235 MHz images. The sensitivities of latter observations are not adequate to detect this candidate relic. In Figure~\ref{rfidtest_zaxisbridge} we notice a {\it bridge} of radio emission between the central halo and the south-west relic in 610 MHz image. This is one of the rare examples  where a radio {\it bridge} is clearly visible between the halo and the relic. Due to lack of sensitivity, this {\it bridge} is not visible in the 150 and 235 MHz observations. 
\par We estimate the equipartition magnetic fields in the MACSJ1131.8-1955 radio halo and (south-west) relic to be  $\sim$ 1 $\mu$G and $\sim$ 1.3 $\mu$G, respectively.   

\subsection{MACSJ0308.9+2645}
\par
\begin{figure*}[t]
    \centering
    \begin{subfigure}[t]{0.38\textwidth}
        \includegraphics[width=\textwidth]{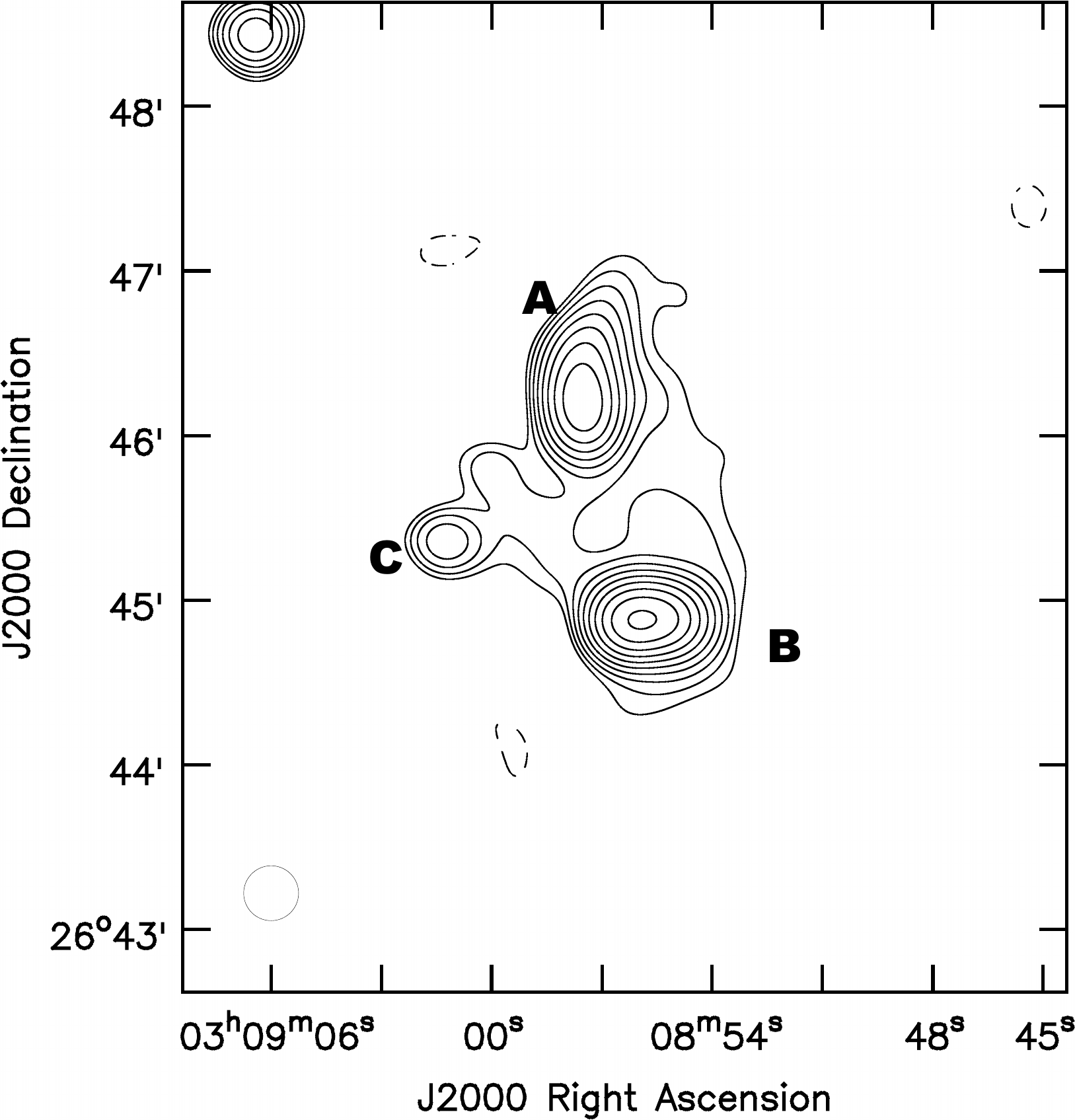}
        \caption{}
        \label{rfidtest_xaxis}
    \end{subfigure}
    \begin{subfigure}[t]{0.38\textwidth}
        \includegraphics[width=\textwidth]{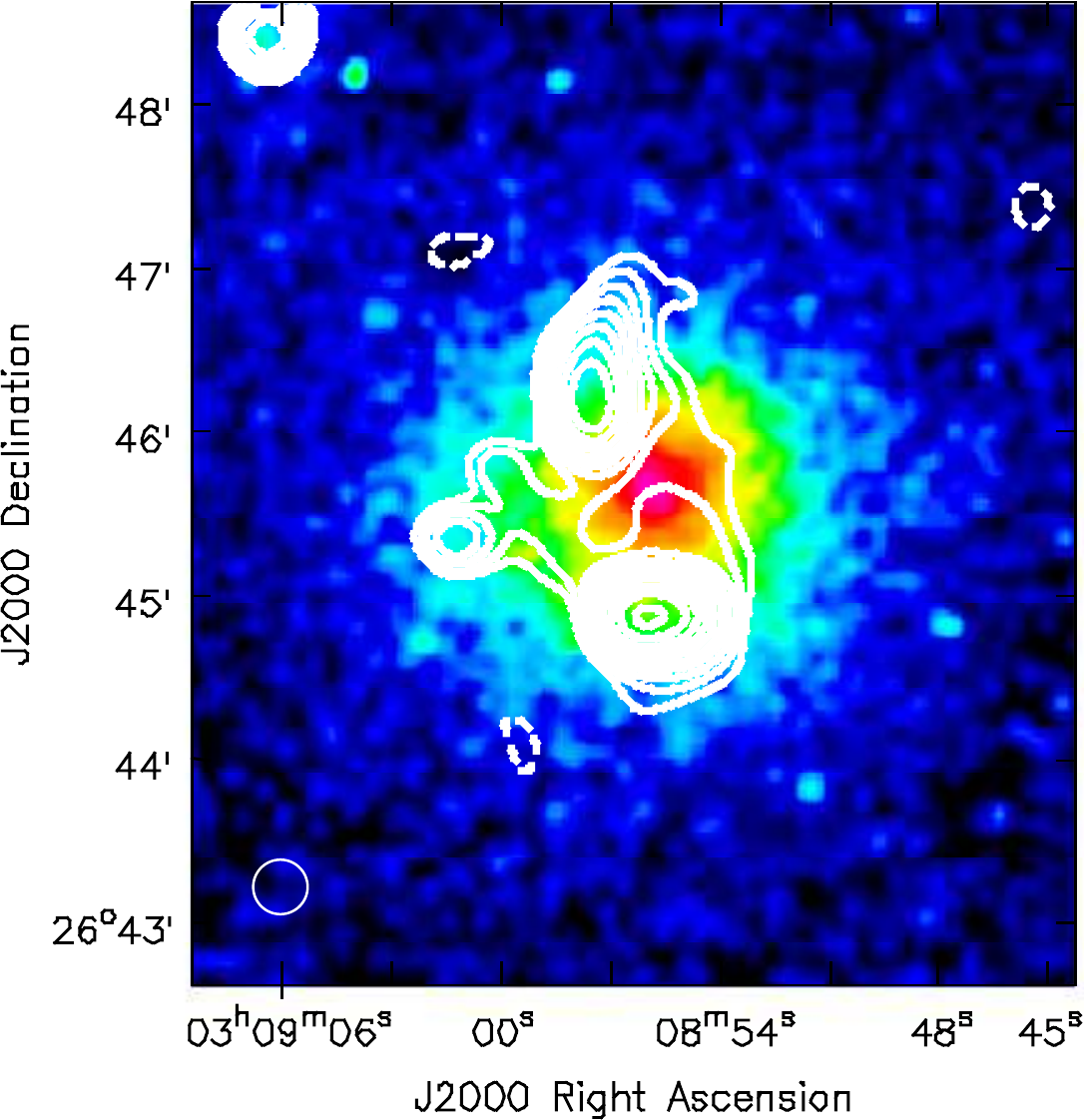}
        \caption{}
        \label{rfidtest_yaxis}
    \end{subfigure}
    \caption[]{ MACSJ0308.9+2645. (a) GMRT 610 MHz (resolution of 20$''$ $\times$ 20$''$) contours, and (b) GMRT 610 MHz contours on {\it Chandra} X-ray image. First contour is drawn at 3$\sigma$ where $\sigma_{610 MHz}$ = 0.38 mJy beam$^{-1}$. Contour level increases in steps of $\sqrt{2}$. Dashed line shows -3$\sigma$ contours.}
    \label{MACSJ0308_radio_img}
\end{figure*}

 MACSJ0308.9+2645 is a massive (10.75 $\times$ 10$^{14}$ $M_{\odot}$), luminous (1.79 $\times$ 10$^{45}$ erg s$^{-1}$), and hot (10.54 keV) cluster situated at $z$ = 0.35. We analysed 150, 235, and 610 MHz GMRT data to search for diffuse radio emission in this cluster. We detected candidate radio halo, only in 610 MHz observation. Figure~\ref{MACSJ0308_radio_img} shows (a) 610 MHz image, and (b) 610 MHz contours overlaid on the {\it Chandra} X-ray image. Its X-ray morphology seems to be relaxed and regular. However, its temperature distribution shows two arc like substructures (see \S \ref{discn}) and radio halo is situated between these substructures. The largest linear size of the candidate radio halo is $\sim$ 850 kpc. There are three point sources (A, B and C) embedded in the diffuse radio source. We report 610 MHz integrated flux density (after subtracting flux densities of point sources) of the halo within 3$\sigma$ contour level in Table \ref{MAC_clusters_rad_obs}. There is no diffuse radio emission visible in the 150 and 235 MHz images. The expected surface brightness of the radio halo is $\sim$ 3.3 mJy beam$^{-1}$ at 150 MHz, and $\sim$ 1.8 mJy beam$^{-1}$ at 235 MHz. Both these expected values are below the rms value ($\sim$ 4 mJy beam$^{-1}$) in these frequencies. We estimated the radio power at 1400 MHz ($P_{1.4GHz}$) by extrapolating the flux density at 610 MHz to flux density at 1400 MHz using a spectral index of -1.3 (which is the average value of spectral index of sample of halos given in \cite{2012A&ARv..20...54F}). The equipartition magnetic field in the MACSJ0308.9+2645 radio halo is $\sim$ 0.7 $\mu$G.

\subsection{MACSJ2243.3-0935}
\par This cluster is located at the centre of supercluster of SCL2243-0935, at a redshift of $z$ = 0.44. Three massive filaments of size (5-15) $h^{-1}_{70}$ Mpc and mass of 1.53 $\pm$ 1.01 $\times$ 10$^{15}$ $M_{\odot}$ have been discovered around this cluster. The size of this supercluster is estimated to be 45 $\times$ 15 $\times$ 50 $h^{-1}_{70}$ Mpc ($l \times w \times d$), making it one of the largest supercluster known at intermediate redshifts \citep{2011A&A...532A..57S}. There are $\sim$ 14 cluster members embedded in this large filamentary cosmic web network. As compared to other superclusters such as A901/902, MS0302 or CL0016, SCL2243-0935 is significantly larger in size, mass and host member clusters. The core cluster of SCL2243-0935, MACSJ2243.3-0935 is massive ($M_{200}$ = 1.54 $\pm$ 0.29 $\times$ 10$^{15}$ $M_{\odot}$) and shows significant sub-structuring. It contains several strongly lensed galaxies and it is not in dynamical equilibrium based on galaxies distribution \citep{2011A&A...532A..57S}.    
\par
\begin{figure*}[t]
    \centering
    \begin{subfigure}[t]{0.38\textwidth}
        \includegraphics[width=\textwidth]{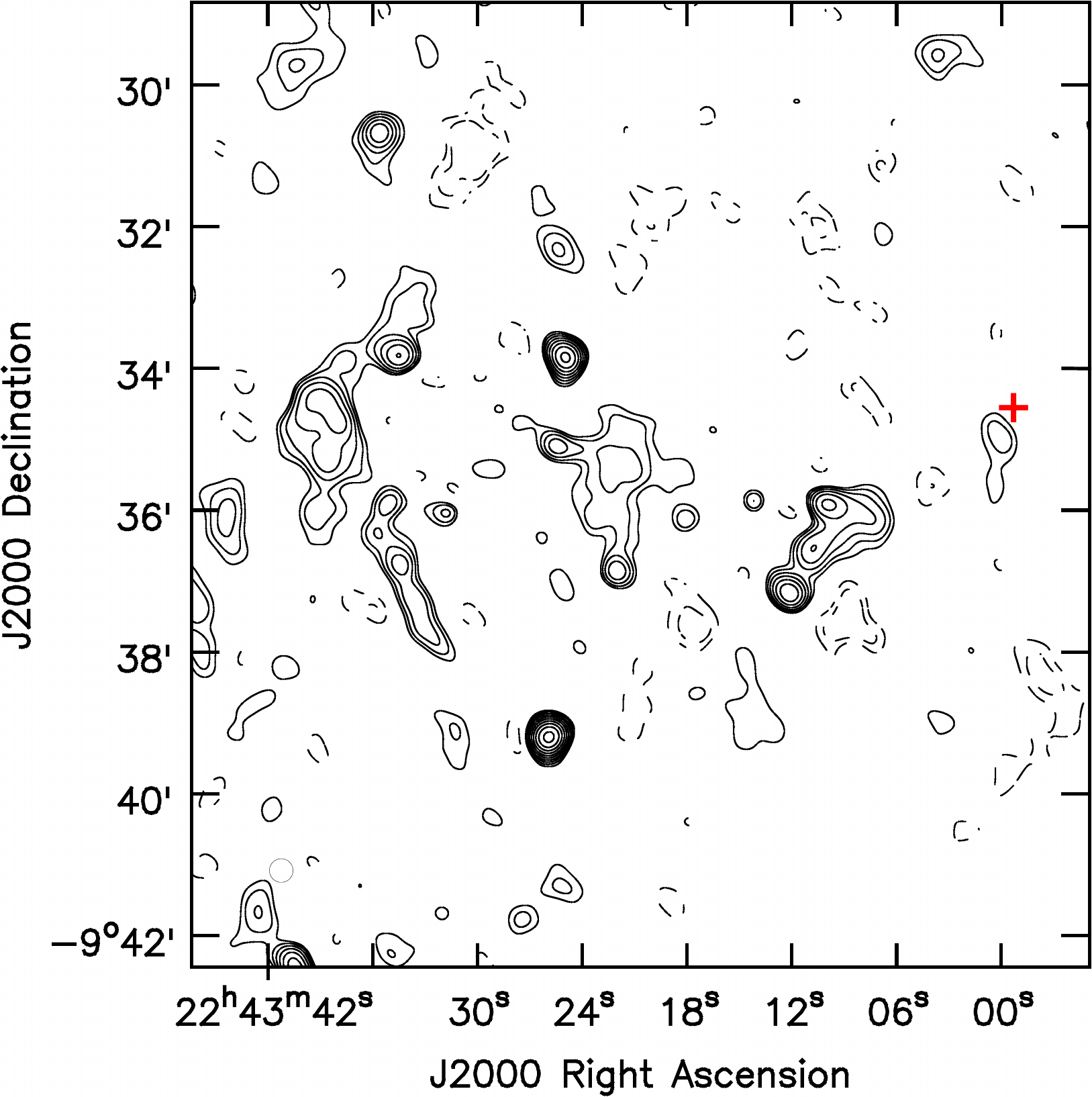}
        \caption{}
        \label{rfidtest_xaxis}
    \end{subfigure}
    \begin{subfigure}[t]{0.38\textwidth}
        \includegraphics[width=\textwidth]{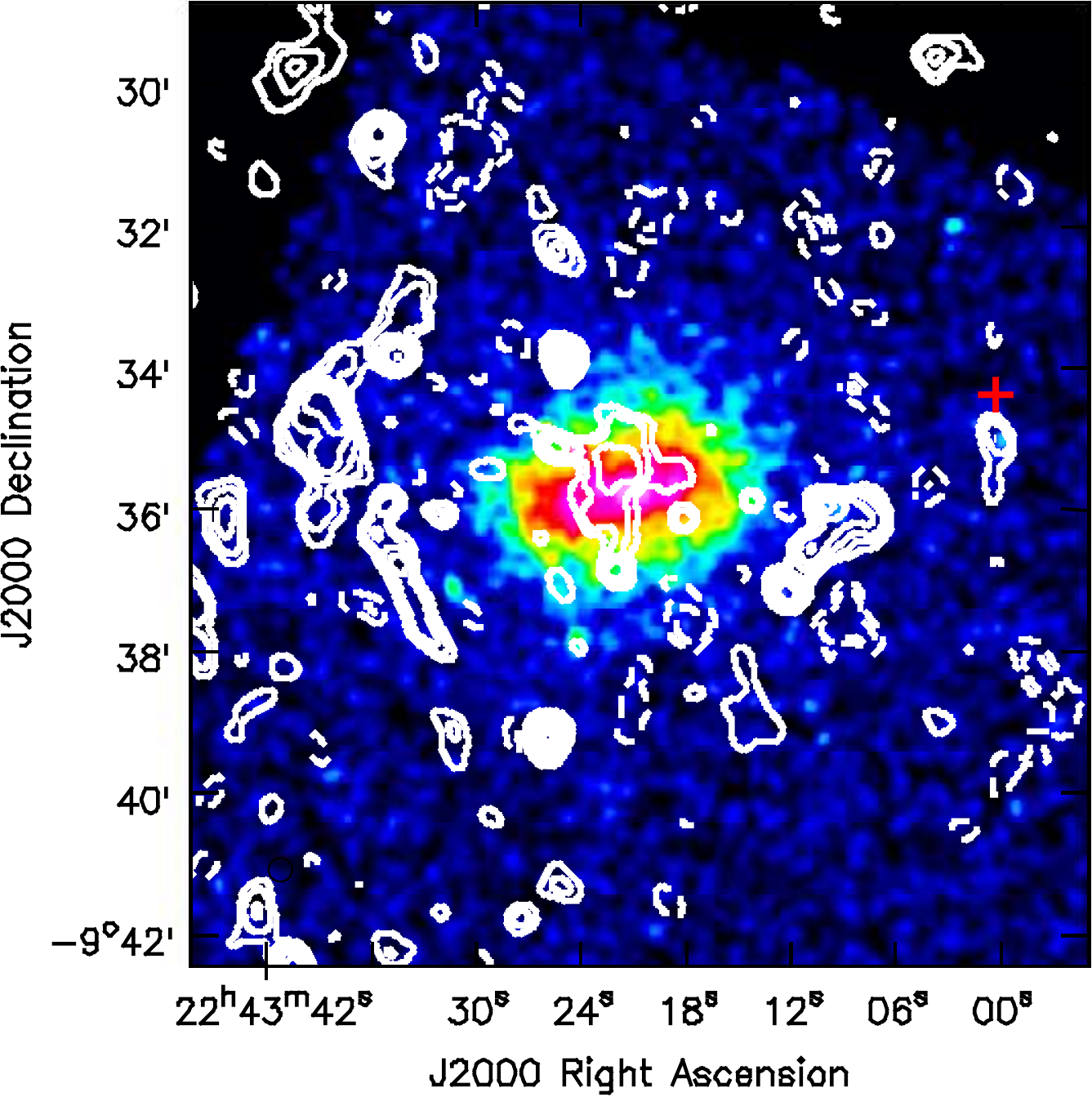}
        \caption{}
        \label{rfidtest_yaxis}
    \end{subfigure}
    \caption[]{ MACSJ2243.3-0935. (a) GMRT 610 MHz contours (resolution of 20$''$ $\times$ 20$''$) for a larger region around the core of cluster, and (b) GMRT 610 MHz contours on {\it Chandra} X-ray image for the central halo region. First contour is drawn at 3$\sigma$ where $\sigma_{610 MHz}$ = 0.10 mJy beam$^{-1}$. Contour level increases in steps of $\sqrt{2}$. {\color{red} +}  sign are for the candidate relic. Dashed line shows -3$\sigma$ contours.}
    \label{MACSJ2243_radio_img}
\end{figure*}

 We analysed 235 \& 610 MHz GMRT data, and 1400 MHz EVLA data to search for diffuse radio emission in MACSJ2243.3-0935. No diffuse radio emission is visible in the 150 (TGSS), 235 and 1400 MHz data. Figure~\ref{MACSJ2243_radio_img} shows (a) 610 MHz image for larger region around the core of the cluster, and (b) 610 MHz contours on {\it Chandra} X-ray image for the central core region of the cluster. We detected candidate radio halo of largest linear size of $\sim$ 722 kpc, and perpendicular to the merger axis in 610 MHz image. X-ray observations clearly show a disturb morphology. There are many active radio galaxies visible around the cluster mainly in east direction from the cluster centre. Some of them have one sided or two sided jet emission. Some of them have optical counterparts in the SDSS survey. Recently, \cite{2016arXiv160205923C} have detected a candidate relic in MACSJ2243.3-0935 using GMRT 610 MHz newer observation. This candidate relic is marked with `{\color{red}+}' sign in Figure~\ref{MACSJ2243_radio_img}. This relic is situated in west direction at distance of 2.13 Mpc from the cluster centre. We estimated largest linear size of it $\sim$ 680 kpc which is same as \cite {2016arXiv160205923C}. However our flux density measurements of 2.0 $\pm$ 0.4 mJy is not  consistent with their measurements. More details on discrete radio sources around the radio halo are given in \cite {2016arXiv160205923C}. 
In the X-ray image (Figure~\ref{MACSJ2243_radio_img}(b)), there are two subclusters in the west and east direction, and diffuse radio emission is situated in between them. As shown in Figure 4 of \cite{2011A&A...532A..57S} MACSJ2243.3-0935 has two clumps of galaxy distribution. This suggests that there are two clusters which are merging or have already passed through each other (post merger stage). 
\par We report 610 MHz integrated flux density of the halo in MACSJ2243.3-0935 measured within 3$\sigma$ contour level in Table \ref{MAC_clusters_rad_obs}. We estimated  expected surface brightness values for the halo at 150 ($\sim$ 2.11 mJy beam$^{-1}$), 230 ($\sim$ 1.18 mJy beam$^{-1}$), and 1400 MHz ($\sim$ 0.11 mJy beam$^{-1}$). These expected values are at the level of 2-3 $\sigma$ in these observations. We estimated the radio power at 1400 MHz ($P_{1.4GHz}$) by extrapolating the flux density at 610 MHz to flux density at 1400 MHz using spectral index of -1.3. The equipartition magnetic field in this candidate radio halo is $\sim$ 0.7 $\mu$G. 

\subsection{MACSJ2228.5+2036 and MACSJ0358.8-2955}
\par These two MACS clusters are massive and hot as well as disturbed in their X-ray morphology, but there is no diffuse radio emission visible at any of the observed frequencies (Table \ref{MAC_clusters_rad_obs}). \cite{2008A&A...484..327V} have first observed MACSJ2228.5+2036 using GMRT as a part of GMRT radio halo survey. They have also not detected any diffuse radio emission in it. Hence we calculate the upper limit on the radio luminosity of the halos that could be presence in these two clusters. We report upper limits for these two clusters in Table \ref{MAC_clusters_rad_obs}.  
We calculated the upper limit (at 610 MHz) by first measuring the rms in the central regions of these two clusters. This rms is measured for a single beam, so we need to scale it with a {factor of $\sqrt{(size^{2}/beam~area)}$} to take an account for number of beams within the halo size. {We used a canonical size for the radio halos to be $\sim$ 1 Mpc}. An upper limit to the radio power of a halo at 1400 MHz can be estimated using the upper limit at 610 MHz. We used a spectral index of -1.3 to extrapolate upper limit at 610 MHz to 1400 MHz.

\begin{table*}
\centering
\caption{Radio observations.}
\label{MAC_clusters_rad_obs}
\begin{tabular}{cccccccc}
\hline
Cluster &Frequency & angular & linear & flux density &log$P_{1.4GHz}$  \\
& & size & size &  of halo and relic & &  \\
&(MHz) &$''$ $\times$ $''$&Kpc $\times$ Kpc &(mJy)&(W Hz$^{-1}$) \\
\hline
MACSJ0417.5-1154 &235&118$\times$180 &671$\times$1024&      77$\pm$8.0 & \\
&610&130$\times$210 &739$\times$1194&      54$\pm$5.5 & \\
&1575&115$\times$180 &654$\times$1024&    10.60$\pm$1.0 & 24.45 \\
\hline
MACSJ1131.8-1955 (halo)&150  &115$\times$190 &512$\times$847&502$\pm$58 & \\
&235  &115$\times$227 &512$\times$1012&228$\pm$31  &             \\
&325$^\dagger$  &198  &890&130$\pm$7  &             \\
&610  &112$\times$183 &499$\times$815&   70$\pm$7.2 &             \\
&1400$^\ddagger$ &108   &540&20$\pm$2.8$^{\star}$ & 24.51 \\
\hline
MACSJ1131.8-1955 (relic)&150 &50$\times$190 &240$\times$849  &228$\pm$30 &     \\
&235 &60$\times$85 &267$\times$379&  156$\pm$18 &    \\
&325$^\dagger$ & 99.2&450& 75$\pm$4   &    \\
&610 &60$\times$100  &267$\times$446&  46$\pm$5 &    \\
&1400$^\ddagger$ &150 &750&20$\pm$2.3$^{\star}$ & 24.51            \\
\hline 
MACSJ0308.9+2645 &610 &132$\times$172 &652$\times$850&30$\pm$4.13 & 24.32    \\
\hline
MACSJ2243.3-0935&610 &83$\times$127 &472$\times$722 &9$\pm$1.03 & 23.97      \\
\hline
MACSJ2228.5+2036&610 &  & & $<$3.13 & $<$23.44 \\
\hline
MACSJ0358.8-2955&610 & &&$<$6.0 &$<$23.76 \\
\hline
\end{tabular}

\hspace{-9.3cm}$\dagger$ = \protect \cite{2013A&A...551A..24V} and $\ddagger$ = \protect \cite{1999MNRAS.302..571R}. $^{\star}$ = NVSS values.  
\end{table*}

\begin{table*}
\centering
\caption{Flux densities of unresolved sources in MACS clusters. }
\label{unresoved_src}
\begin{tabular}{cccccccc}
\hline
Cluster &Frequency & flux density   \\
& &  of unresolved point sources   \\
&(MHz) &(mJy) \\
\hline
MACSJ0417.5-1154 &235& 46$\pm$4.6 17$\pm$1.7 (A \& B) \\
&610& 30$\pm$3 6$\pm$0.6       \\
&1575&13$\pm$1.3 4$\pm$0.4\\
\hline
MACSJ1131.8-1955 (halo)&150  &357$\pm$37 114$\pm$11 (A \& B)  \\
&235  &580$\pm$58 190$\pm$19              \\
&610  &85$\pm$8.5 30$\pm$3                 \\
\hline
MACSJ0308.9+2645 &610 &30$\pm$3 42$\pm$4.2  3$\pm$0.3 (A \& B \& C)  \\
\hline
\end{tabular}
\end{table*}

\section{X-ray data analysis and Images}
\par In this work, we used {\it Chandra} archival data for the sample of six MACS clusters (Table \ref{xraydata}). We processed these data with CIAO 4.6 and CALDB 4.6.1.1. We first used the \verb"chandra_repro" task to reprocess all ACIS imaging data, removed any high background flares (3$\sigma$ clipping) with the task \verb"lc_sigma_clip", followed by combining multiple data sets using \verb"merge_obs" script. All event files included the 0.3--7 keV broad energy band and 2$''$ pixels binning. We removed point sources around the cluster. We divided counts image with exposure map and generated the flux image (photon / cm$^{2}$ / s). Finally we smoothed this image with $\sigma$ = 10$''$ to remove zero counts. 

\begin{table*}
\centering
\caption{MACS clusters {\it Chandra} X-ray data.}
\label{xraydata}
\begin{tabular}{ccccc}

\hline
Cluster                &Obsid                     &Total exposure time (ks) \\
\hline
MACSJ0417.5-1154  &    11759, 12010, 3270      &             90  \\
MACSJ1131.8-1955  &      15300, 3276           &             24 \\
MACSJ0308.9+2645  &    3268                    &             23 \\
MACSJ2243.3-0935  &     3260                   &             20 \\
MACSJ2228.5+2036  &     3285                   &             20 \\
MACSJ0358.8-2955  &    11719, 12300, 13194     &             60 \\
\hline 
\end{tabular} 

\end{table*}

\subsection{Temperature maps}
\par To understand the hot ICM morphology and its relation with diffuse radio sources, we generated temperature maps of MACS clusters. The temperature map of the ICM is also a good indicator of the dynamical activity in the cluster. We do not show temperature maps for the non-detection radio halo clusters i.e. MACSJ2228.5+2036 and MACSJ0358.8-2955. For the remaining clusters, we used the highest exposure time data. We used XMC (X-ray Monte Carlo) technique to generate temperature map \cite[see][for details of the method]{2007ApJ...655..109P, 2007ApJ...670.1010A, 2009ApJ...696.1029A}. In this technique, it is assumed that the given intra-cluster hot gas is composed of a number of Gaussian ``blobs'' or smooth particles. Each blob has its own temperature, flux, abundance, electron density, phi and psi (sky location), Gaussian width (size), redshift, absorption column, etc. X-ray emission from these hypothetical ``blobs'' is then propagated through the (simulated) telescope's instrument, generating dummy X-ray events in a similar way to that which would come from an actual X-ray observation. These simulated events are compared with the real data events being modelled. In this method the spectral and spatial models are used together. We used the warm-absorbed APEC (spectral) model along with the {\it Chandra} detector (spatial) model derived for the XMC. In this procedure, given (free) parameters (APEC model parameters such as temperature and normalisation, and spatial coordinates of detectors) of the ICM are iterated using the MCMC { (Markov Chain Monte Carlo)} technique. Table \ref{xmc_parameter_range} shows the parameters and their ranges. For instance, in this analysis we allowed temperature to vary between $\sim$ 1 and 15 keV, while fixed the solar abundance to 0.3 for all clusters. We used the NRAO H1 all sky survey data \citep{1990ARA&A..28..215D} to derive the nH values for each cluster. 
\par
\begin{figure*}
\centering
\hspace*{-0.3in}
\includegraphics[scale=0.52]{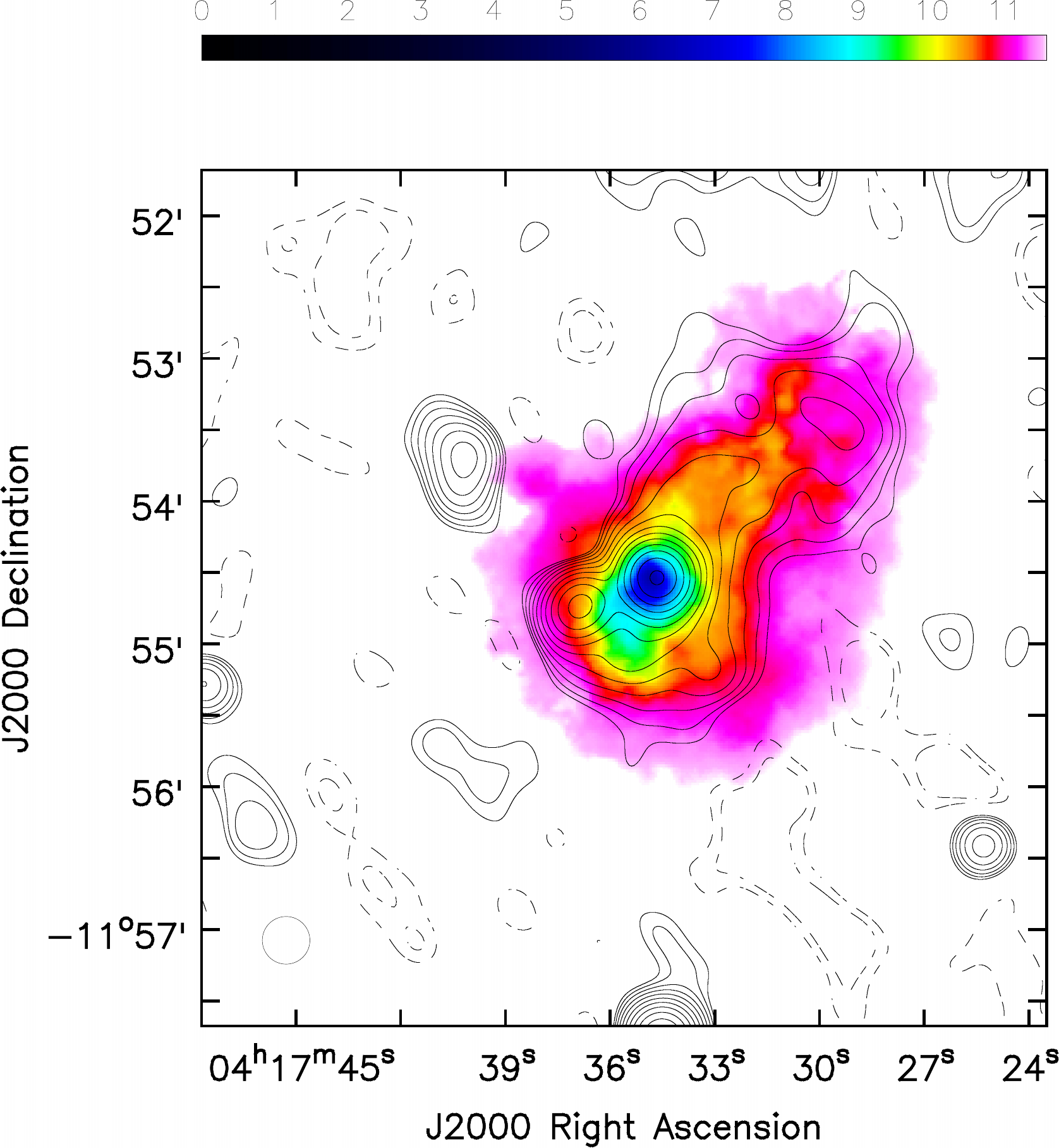}
\includegraphics[scale=0.52]{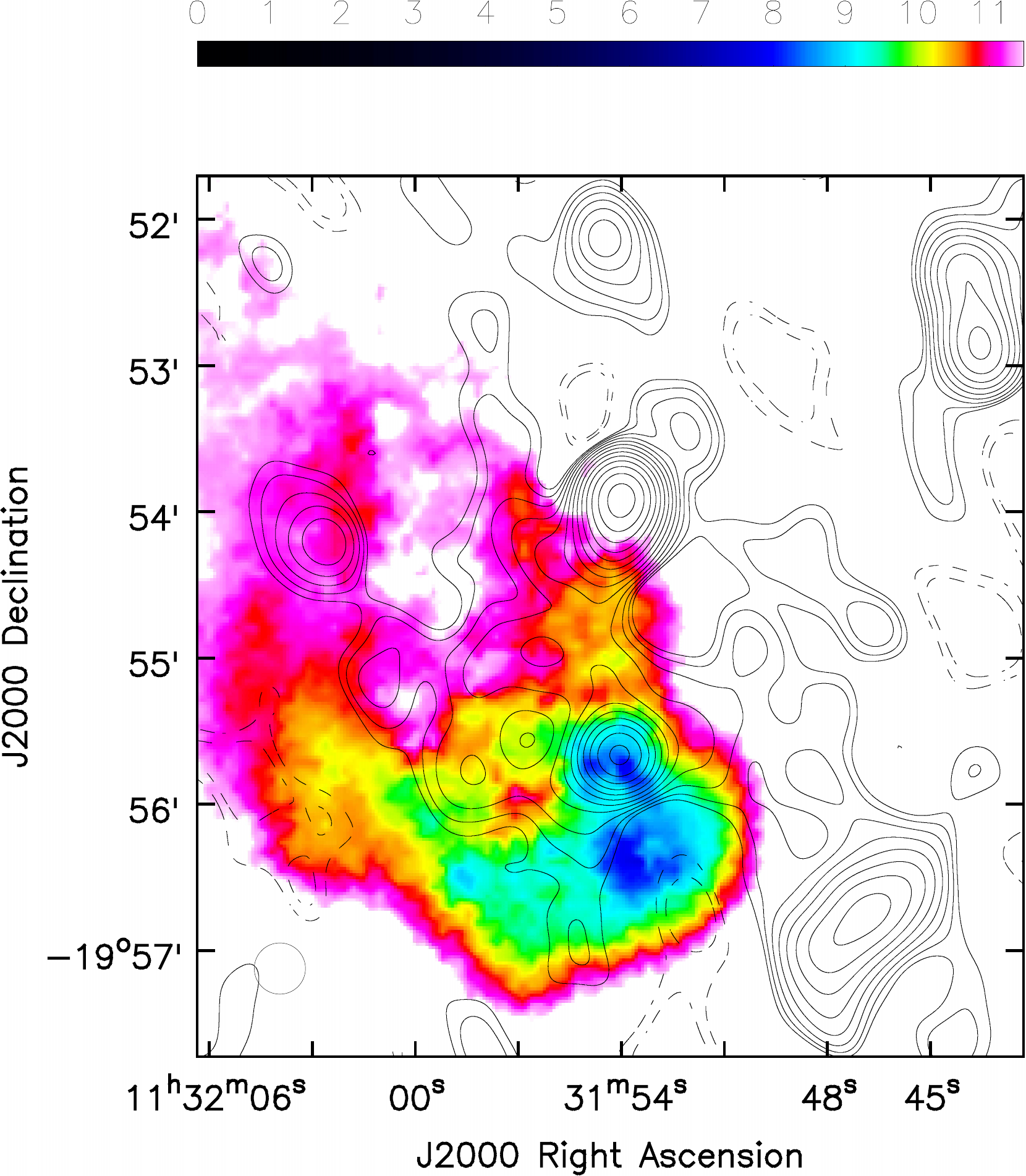}
\caption {(Left) MACSJ0417.5-1154 and (Right) MACSJ1131.8-1955 temperature maps. In both these images, contours are corresponding to 610 MHz GMRT radio observations. Contour levels are the same as the Figure~\ref{MAC0417_radio_img} and \ref{MACSJ1131_radio_img}, respectively. Colorbar is in units of keV. {We displayed temperature values at $\sim$ 30\% confidence level}.}
\label{MACSJ0417_temp}
\end{figure*}

\begin{figure*}
\centering
\hspace*{-0.3in}
\includegraphics[scale=0.50]{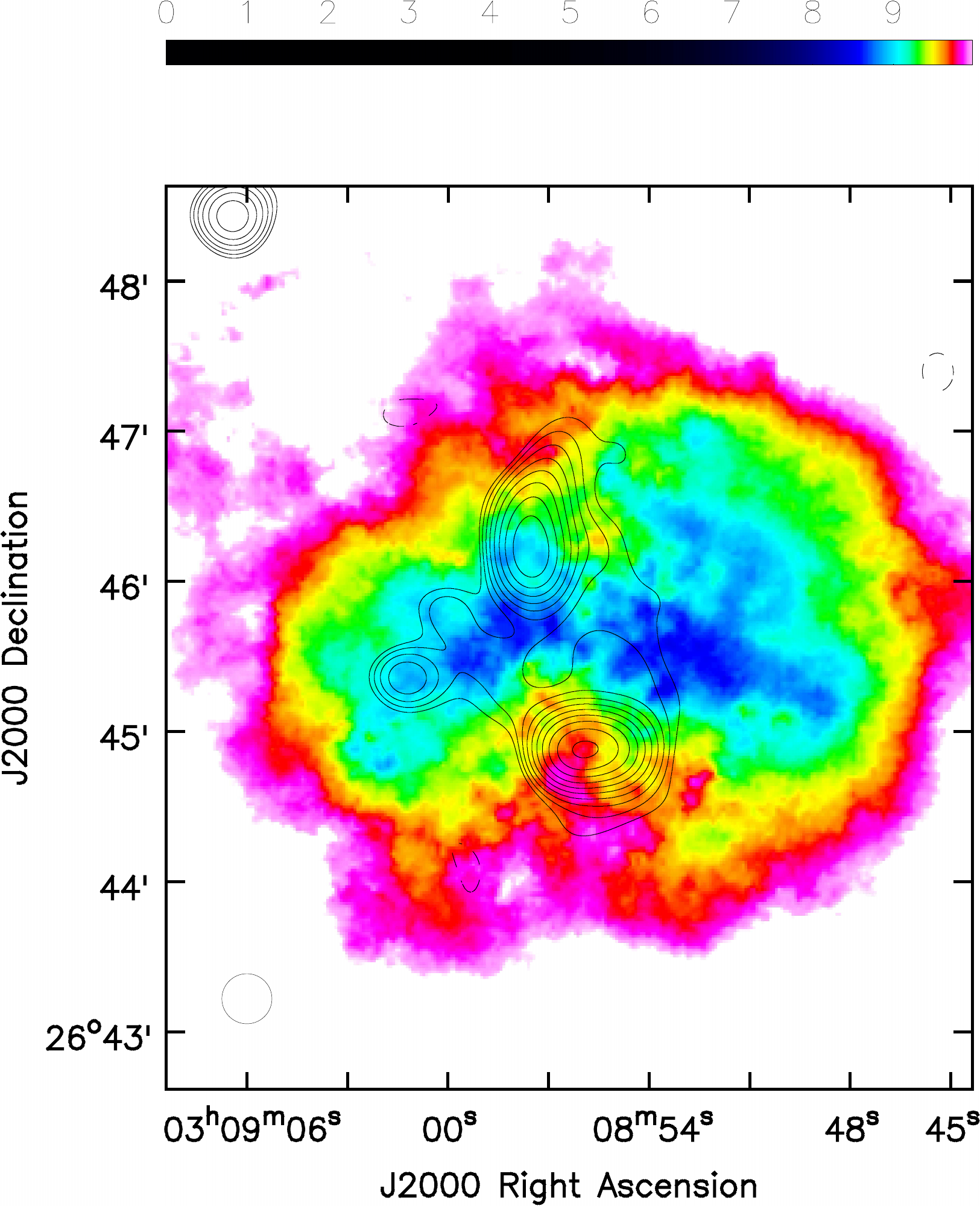}
\includegraphics[scale=0.50]{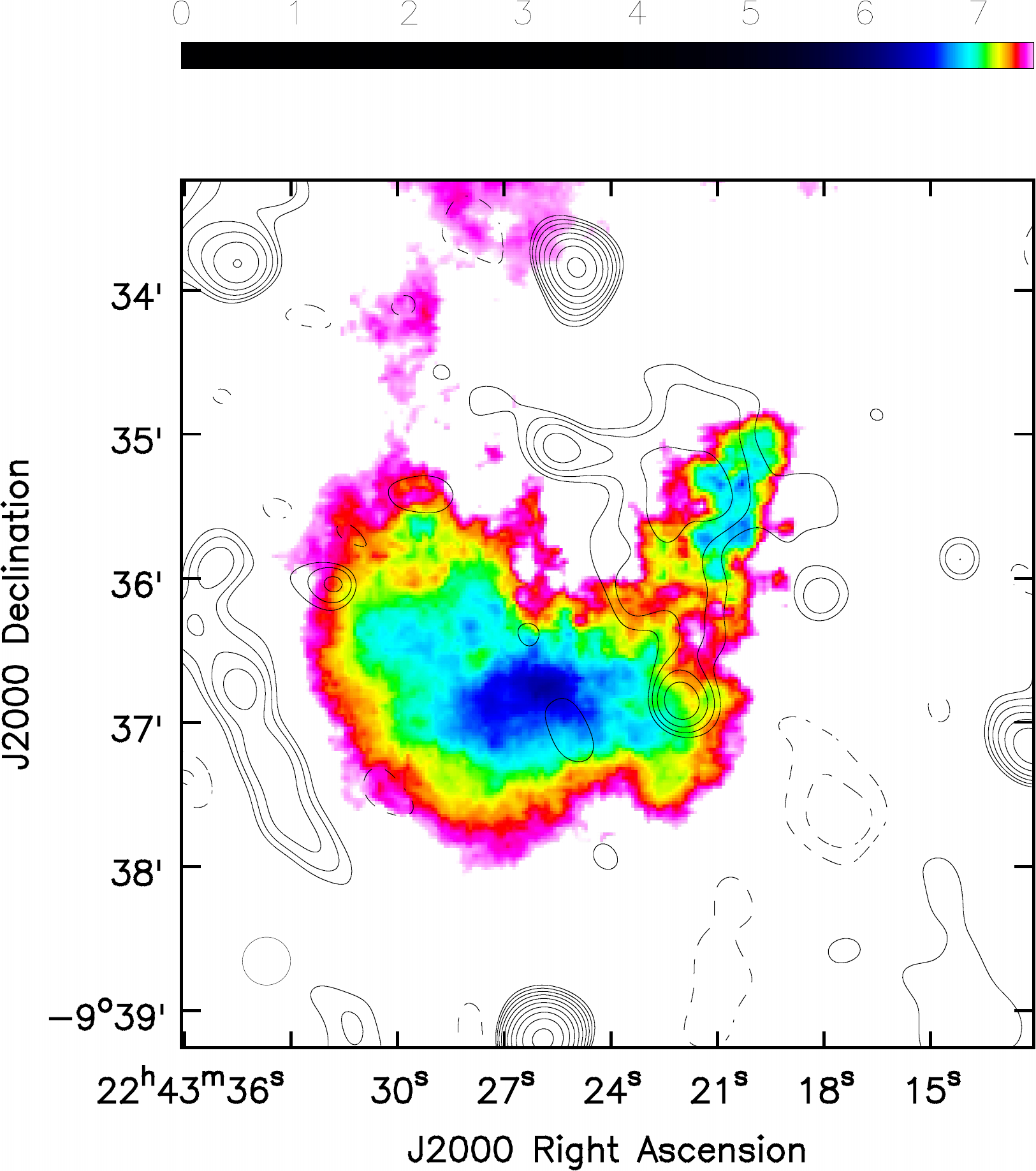}
\caption {(Left) MACSJ0308.9+2645  and (Right) MACSJ2243.3-0935 (for only central halo region) temperature maps. In both these images, contours are corresponding to 610 MHz GMRT radio observations. Contour levels are same as Figure~\ref{MACSJ0308_radio_img} and Figure~\ref{MACSJ2243_radio_img}. Colorbar is in units of keV. {We displayed temperature values at $\sim$ 30\% confidence level}.}
\label{MACSJ0308_temp}
\end{figure*}

 When all parameters converge after a sufficient number of iterations, the resulting distribution of blobs describes the shape and characteristics of the galaxy cluster under investigation \citep{2007ApJ...670.1010A}. The final results have a  number of statistical samples of acceptable or converged fits which fall into the ``confidence region'' of the ICM's input parameters. These well ``fitted'' parameters describe the properties of the ICM. In this work, all the results derived from the model samples are from the iterations from 500 to 3000, where the value of $\chi^{2}$ is reduced (to $\sim$ 1) and stable, and is considered to be a converged chain where all corresponding  parameters have the best-fit value. Figure~\ref{MACSJ0417_temp} and \ref{MACSJ0308_temp} show the temperature maps of four MACS clusters. Temperature map shows high value because in the XMC method multi-temperature plasma overlaps on each other  which mimics the `real' situation in ICM plasma \citep{2009ApJ...696.1029A}. The advantage of the XMC technique is that it explores, with the MCMC method, the extremely high dimensionality of parameter space of the blobs for a given ICM model. The MCMC technique efficiently samples the probability distribution of a model's parameters without being trapped in local likelihood maxima. The result is a well fitted list of sample cluster models, all of which are consistent with the data.

\begin{table*}
 \centering
 \caption{Spectral and Spatial model parameters and their values for XMC analysis.}
 \label{xmc_parameter_range}
 \begin{tabular}{ccc}
  \hline
 
  Parameters & Fixed/Free & min and max values \\ 
  \hline
  Spectral model &  &  \\ 
  $n_{H}$ & Fixed & NRAO H1 all sky survey \\ 
  T (keV) & Free & 1 - 15keV \\ 
  $Z_{\odot}$ & Fixed &  0.3 \\ 
  z & Fixed &  According to Table \ref{MAC_clus_prop}\\ 
  Spatial model &  &  \\ 
  x & Free & -8.4$'$ - 8.4$'$ \\ 
  y & Free & -8.4$'$ - 8.4$'$ \\ 
  ln $\sigma$ ($''$)& Free &  0 - 6$''$\\ 
  \hline 
 \end{tabular} 

\end{table*}

\subsection{X-ray morphology analysis}
\par Substructures, X-ray morphology, and/or X-ray centroid variations are typical features that allow us to understand whether or not a cluster is virialized. The X-ray emitting gas in galaxy clusters carries signatures of dynamical activity that manifests itself as distortions in the X-ray surface brightness images. We computed three non-parametric morphology parameters, $Gini$, $M_{20}$ and Concentration ($C$)  using the {\it Chandra} X-ray images of MACS clusters to characterise the degree of disturbances in its ICM. These three parameters are known to be effective in segregating galaxy clusters according to the level of disturbances in them (P15). $Gini$ is a measure of the flux distribution among the image pixels; its value is 0 if the flux is equally distributed among the pixels (typically non-relaxed clusters) and is 1 if most of the flux is contained only in a small number of pixels (typically relaxed and cool-core clusters) \citep{2004AJ....128..163L}. The moment of light, $M_{20}$, is the normalised second order moment of the relative contribution of the brightest $20\%$ of the pixels \citep{2004AJ....128..163L} and is a measure of the spatial distribution of bright cores and substructures in the cluster. The typical values of $M_{20}$ are  
between -2.5 (very relaxed) to -0.7 (very disturbed) (see P15). The parameter $C$ is a measure of the concentration of the flux in the cluster that depends on the ratio of the radii at which $80\%$ and $20\%$ of the cluster flux is found \citep{2003ApJS..147....1C} and has a minimum value of 0.0 which indicates the most disturbed clusters.  
\par
\begin{figure}
 \centering
 \includegraphics[width=\columnwidth]{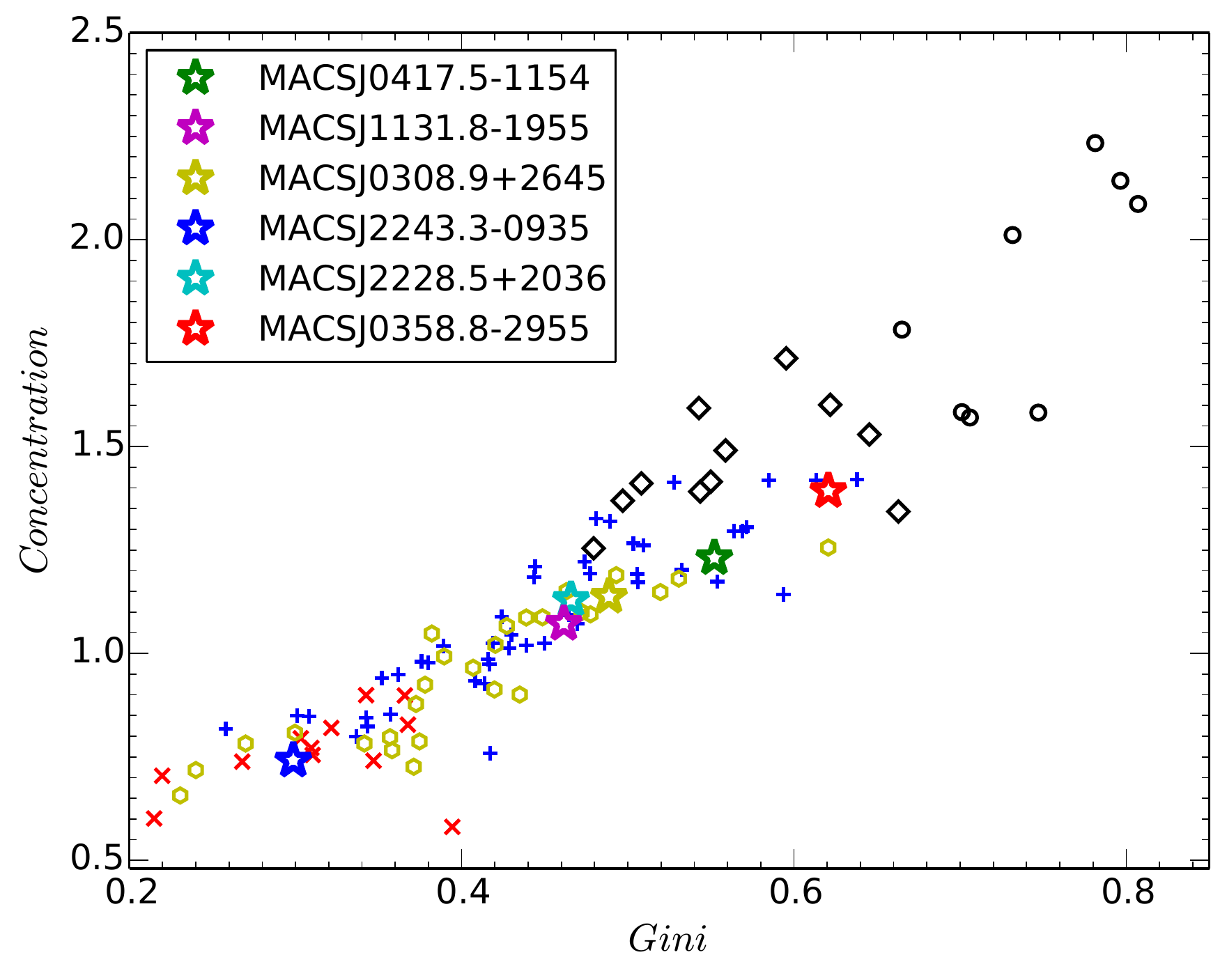}
 \includegraphics[width=\columnwidth]{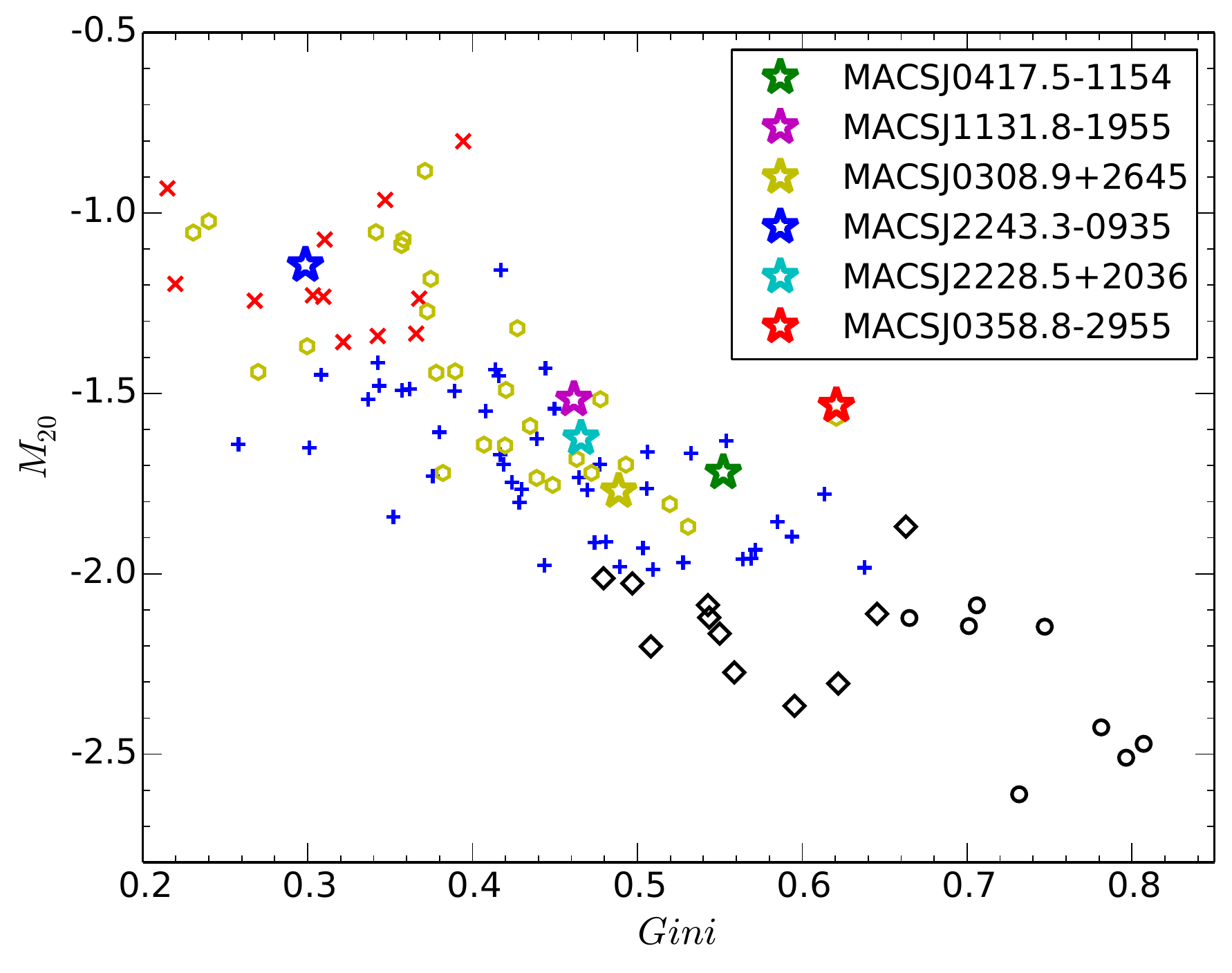}
 \includegraphics[width=\columnwidth]{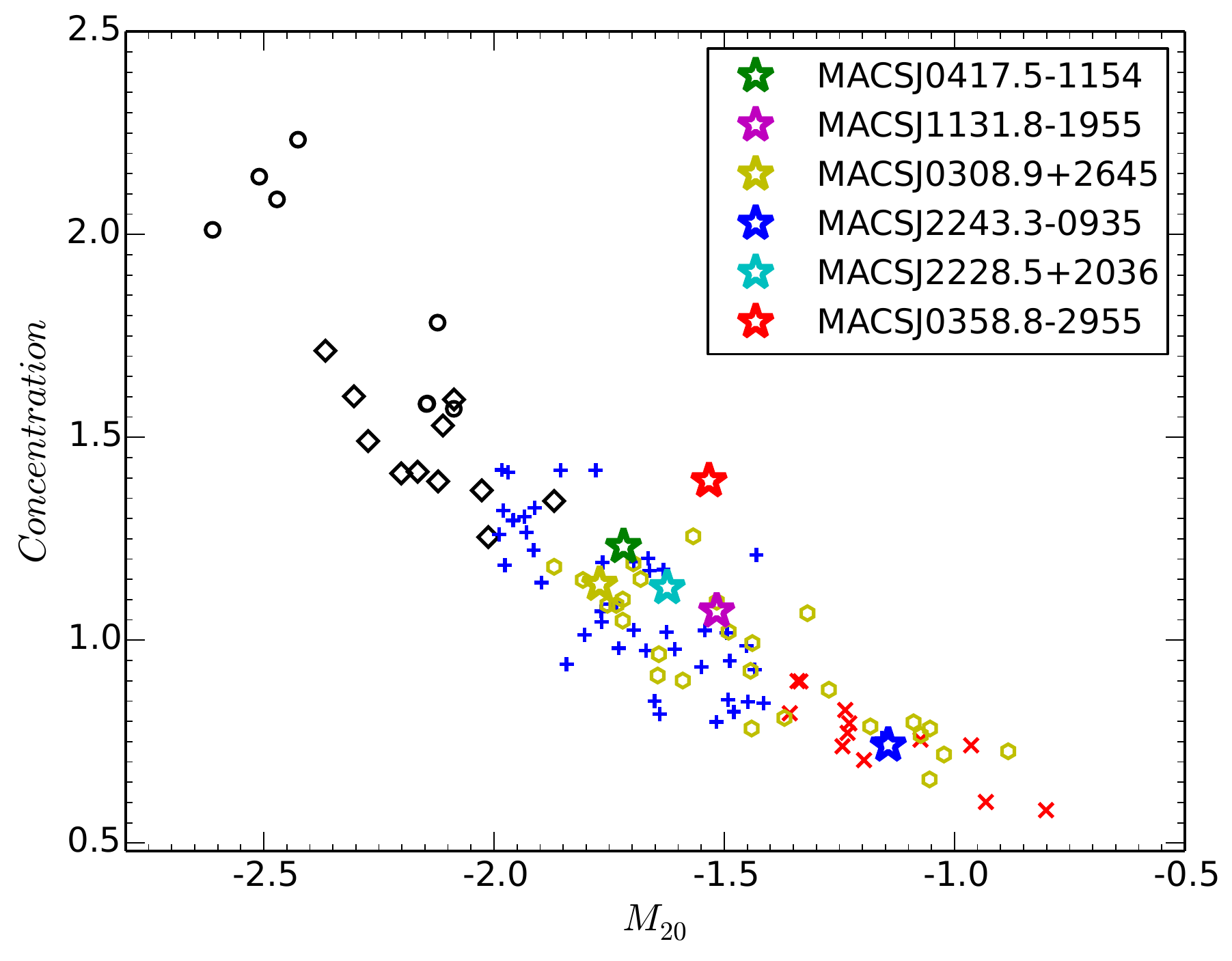}
 \caption {\small Sample of MACS clusters are shown in morphology 
parameter planes along with the P15 sample of clusters. The P15 sample clusters are shown according to their classification --  circles are `strong relaxed', diamonds are `relaxed', plus signs are `non-relaxed' and crosses are `strong non-relaxed' clusters. The hexagons are clusters that host radio halos from \citep{2009A&A...507.1257G} and known to be merging clusters. The clusters in the current sample are marked as different color stars: MACSJ0417.5-1154 (green), MACSJ1131.8-1955 (magenta), MACSJ0308.9+2645 (yellow), MACSJ2243.3-0935 (blue), MACSJ2228.5+2036 (cyan), and MACSJ0358.8-2955 (red).}
 \label{morph_para_plt}
\end{figure}

\begin{figure*}[htb]
\centering
\hspace*{-0.8in}
\includegraphics[scale=0.35]{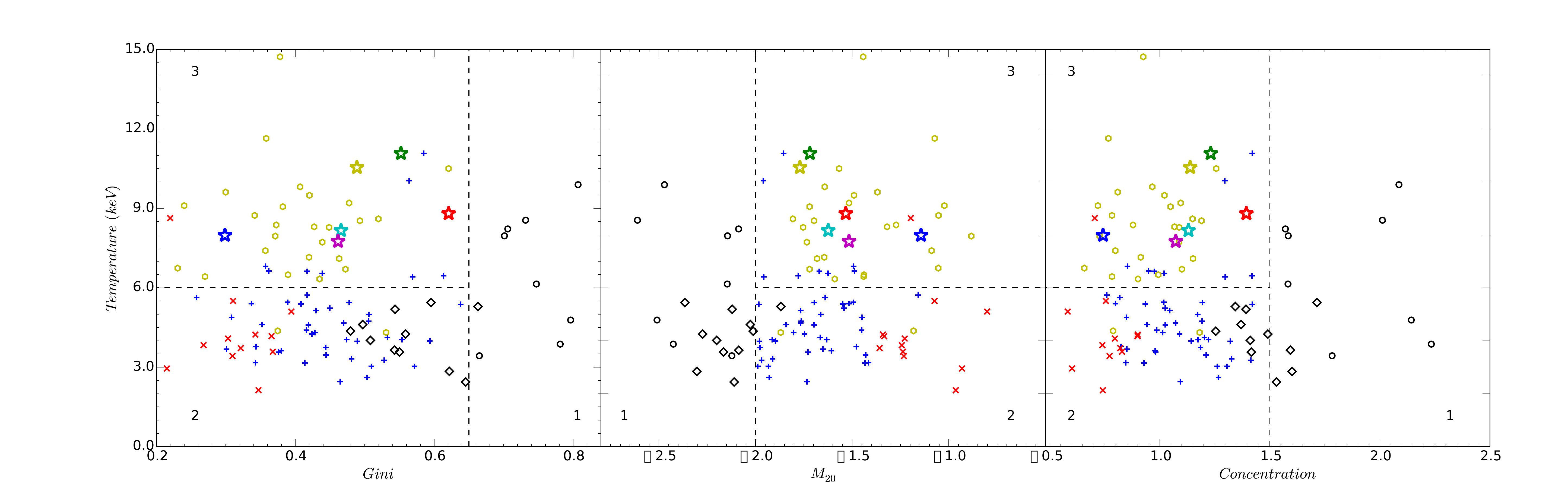}
\caption {\small Morphology parameters vs. Temperature for the sample clusters as in Figure~\ref{morph_para_plt}. We subdivided the parameter vs temperature plot into three regions: (1) has all dynamically relaxed clusters, (2) has radio quiet merger clusters, and (3) has radio loud or radio halo merger clusters.} 
\label{morph_para_temp}
\end{figure*}

 For these MACS clusters we used 500 kpc region around the cluster centroid to calculate the morphology parameters. We have listed the morphology parameters in Table \ref{morph_val} along with 1$\sigma$ uncertainties. We point the reader to P15 and the references therein for the details of the calculations of X-ray morphology parameters. We have plotted each of this parameter vs. other two in Figure~\ref{morph_para_plt}. In this figure, we compared the dynamical state of six MACS clusters with P15 sample clusters. We also plotted morphology parameters vs. cluster temperature in Figure~\ref{morph_para_temp}. In this plot, we compared position of six MACS clusters with other radio halo clusters \citep{2009A&A...507.1257G}. We subdivided this plot into three regions: region (1) has all dynamically relaxed clusters, region (2) has radio quiet merger clusters, and region (3) has radio loud merger (or radio halo) clusters which have temperature $>$ 6 keV.      

\begin{table*}
\centering
\caption{MACS clusters morphology parameter values.}
\label{morph_val}
\begin{tabular}{cccccccc}
\hline
Cluster &$Gini$ & $M_{20}$ & $Concentration$ &  \\
\hline
MACSJ0417.5-1154 & 0.55 $\pm$ 0.01 & -1.72 $\pm$ 0.34 & 1.23 $\pm$ 0.18 \\
MACSJ1131.8-1955 & 0.46 $\pm$ 0.00 & -1.52 $\pm$ 0.30 & 1.07 $\pm$ 0.37 \\
MACSJ0308.9+2645 & 0.49 $\pm$ 0.00 & -1.77 $\pm$ 0.29 & 1.14 $\pm$ 0.44 \\
MACSJ2243.3-0935 & 0.30 $\pm$ 0.00 & -1.14 $\pm$ 0.22 & 0.74 $\pm$ 0.39 \\
MACSJ2228.5+2036 & 0.47 $\pm$ 0.00 & -1.62 $\pm$ 0.37 & 1.13 $\pm$ 0.56 \\
MACSJ0358.8-2955 & 0.62 $\pm$ 0.00 & -1.53 $\pm$ 0.34 & 1.39 $\pm$ 0.62 \\
\hline
\end{tabular}

\end{table*}

\section{Substructure analysis}
\par 
\label{discn}
\par As discussed in section 4.2, all the MACS clusters in this sample appear
to be  dynamically disturbed. 
The morphological parameters of these galaxy clusters quantify the degree of 
disturbances in the ICMs of the respective clusters. 
We compare the parameters estimated for MACS clusters with those for a 
sample of 85 clusters with deep Chandra observations for which 
morphological parameters are available in P15. 
Among the sample of 85 clusters, there are 49 low-$z$ (0.2 -- 0.3) and 
36 high-$z$ (0.3 -- 0.8) clusters with X-ray flux, $f > 1.4\times10^{-13}$ erg s$^{-1}$ cm$^{-2}$ 
and $L_{X}> 3 \times 10^{43}$ erg s$^{-1}$. 
P15 have classified the clusters based on the combination of morphological 
parameters into  four categories of their dynamical states: \\
{\bf Strong Relaxed (SR)} \\
$C > 1.55$, $M_{20} < -2$ and $Gini > 0.65$;\\
{\bf Intermediate clusters: Relaxed (R) and Non-Relaxed (NR)} \\
$1.0 < C < 1.55$, $-2.0 < M_{20}< -1.4$ and $0.40 < Gini < 0.65$;\\
{\bf Strong Non-Relaxed (SNR)}\\
$C < 1.0$, $M_{20} > -1.4$ and $Gini < 0.40$.\\ 
\par In Figure~\ref{morph_para_plt}, the points corresponding to the 
six MACS clusters are plotted along with those for the P15 sample of 
clusters (\cite{2009A&A...507.1257G}) in the morphological parameter planes. 
The SR clusters are well separated from SNR clusters, 
while intermediate clusters are found in between them. 
The morphological parameters (P15) for the sample of radio halo clusters 
from \cite{2009A&A...507.1257G} are also shown in Figure~\ref{morph_para_plt} (hexagons). 
The clusters with radio halos are mainly in the NR (67\%) and SNR (33\%) categories, while
the MACS clusters are in the intermediate category. Furthermore, except the MACSJ2243.3-0935 (SNR), other five MACS clusters are NR.  
Figure~\ref{morph_para_temp} shows the distribution of these clusters in the
temperature and morphology parameters planes. Note that all the MACS clusters are in region 3
which mostly contains clusters with radio halos.

\par We have also analysed the temperature maps of MACS clusters. 
It is believed that the distribution of temperature provides crucial 
information about the underlying substructure in the ICM which is 
sometimes masked in the surface brightness images. 
Furthermore, the reason for using the temperature maps is that, 
we see hot regions caused by a currently-propagating or just-passed 
merger shock \citep{2003ApJ...586L..19M,2005ApJ...627..733M}. 
In these shock-heated regions, gas expands adiabatically, 
reaching pressure equilibrium with the surrounding environment after 
the passage of the shock. 
We also see patchy temperature distribution, spiral and whirlpool structures. 
because of large-scale gas motions during the mergers 
causing subsequent mixing of gases at different temperatures. 
This re-distribution of temperature depends on 
whether the merger has disrupted the structure of the cool core, 
and the time elapsed since the merger. 
We have analysed the temperature maps of the MACS clusters 
in our sample along with its radio counterpart. 
Below we will discuss the substructure analysis of the sample of six MACS clusters.  

\par {\it MACSJ0417.5-1154}. AM12 classified MACSJ0417.5-1154 as a binary head-on merger of similar mass. 
Temperature map (Figure~\ref{MACSJ0417_temp} (left)) shows disturbances largely
 in the outer parts of the cluster, while the inner core ($\sim$ 100 kpc) 
does not appear to be affected by the merger. 
The morphology of the temperature distribution is similar to that of the diffuse 
radio emission. 

\par {\it MACSJ1131.8-1955}. It's complex X-ray morphology indicates a series of recent mergers, as well as an ongoing merger \citep{1997A&A...326...34L, 1999MNRAS.302..571R, 2012MNRAS.420.2480Z}. AM12 classified this cluster as the most complex mergers that have recently passed, or are still undergoing multiple mergers along various axes. 
Temperature map of MACSJ1131.8-1955 (Figure~\ref{MACSJ0417_temp} (right)) is irregular and 
shows a complex substructure in the core. 
In the temperature map, we observed large scale gas motions as similar to the whirlpool structure noticed by \cite{2005A&A...442..827F}. 
Hot gas ($\sim$ 10 keV) is extended in north-east directions where most of the radio emission is located.
Recently \cite{2012MNRAS.420.2480Z} noticed a possible shock front (with a Mach number M = 1.20 $\pm$ 0.10) 
in the southern part of the cluster where south-west relic is located. 
We also noticed, in the temperature map, a sharp edge near the position of the
 south-west relic. 
There is no evidence for a shock front near the candidate relic toward the north-west.

\par{\it MACSJ0308.9+2645}. Its X-ray surface brightness distribution is 
regular and smooth, however its morphology parameters indicate that it is a 
non-relaxed (NR) cluster.  The temperature map of MACSJ0308.9+2645 (Figure~\ref{MACSJ0308_temp} (left)) is rather complex 
in conformity with it being an NR cluster. These substructures are not visible in the corresponding surface brightness image (Figure~\ref{MACSJ0308_radio_img}). There are two cool cores visible in the temperature map. 
It is possible that these two cores are part of merging clusters. 
Diffuse radio emission (at 610 MHz) is visible in between these two cool cores. 
Radio emission is extending in the north-south direction which is roughly 
perpendicular to the merger axis along the direction joining the cool cores.
There is some extension of radio emission toward the east direction. 
Overall morphology of  MACSJ0308.9+2645 indicates that it is an ongoing merger.

\par{\it MACSJ2243.3-0935}. It is the most disturbed cluster in this MACS sample. All morphology parameters are falling on the non-relaxed category which indicates strong non-relaxed (SNR) cluster state. In this sample, MACSJ2243.3-0935 is the only cluster which falls in this state. Its position in Figure~\ref{morph_para_temp} is in region 3 i.e. clusters with radio halo region. AM12 classified it as a highly disturbed cluster based on the large separation between X-ray peak or centroid and BCG. Its temperature map (Figure~\ref{MACSJ0308_temp} (right)) is irregular and shows substructures. There are two hot regions, one towards the north-west and another towards the south-east. There is trail of hot gas ($\sim$ 7 keV) connecting these two substructures. These substructures are not visible in the corresponding surface brightness image (Figure~\ref{MACSJ2243_radio_img}).  
In lower substructure, disturbed cool core is visible, while cool core is completely absent in the upper substructure. As suggested previously, this cluster is located in the core of supercluster SCL2243.3-0935, surrounded by filaments and it has two concentrations of galaxies corresponding to the two higher temperature region. We found candidate radio halo at 610 MHz which is situated in between these two substructures. MACSJ2243.3-0935 could be the post-merger cluster.

\par{\it MACSJ2228.5+2036} and {\it MACSJ0358.8-2955}. Based on morphology analysis, both of these clusters are NR clusters however we have not detected diffuse radio emission in these clusters. \cite{2013MNRAS.429..833H} have analysed MACSJ0358.8-2955 cluster in detail using the {\it Chandra}, {\it Hubble space telescope}, and the {\it Keck-I} telescope. It shows a high velocity dispersion ($\sim$ 1440 km s$^{-1}$) and complex merger { of at least three} substructures. Based on X-ray and optical morphology they have argued that MACSJ0358.8-2955 is an ongoing merger between two components. However, in the morphology parameter planes, MACSJ0358.8-2955 is situated at the edge of NR clusters. We have put stringent upper limits on the radio powers from these two clusters which is $\sim$ 11 times below in the $L_{x}$--$P_{1.4GHz}$ plot.

\section{Discussion and Conclusion}

\par There are six high-$z$, massive ($>$ 7 $\times$ 10$^{14}$ $M_{\odot}$) and non-relaxed MACS clusters in our sample. We analysed their GMRT and EVLA data to detect diffuse radio sources. Out of six, we detect diffuse radio sources in four clusters (67\%) while remaining two (33\%) are `off-state' clusters. It is well established that not all massive and merging clusters harbour diffuse radio sources \citep{2011MNRAS.417L...1R}. The typical life time of a radio halo is $\sim$ Gyr \citep{2009A&A...507..661B} which is comparable to cluster merging time scale. These activities of merger temporarily change the X-ray global properties such as the bolometric luminosity \citep{2001ApJ...561..621R}. Turbulence generation and propagation processes take time to switch-on the radio emission. This turbulence generation activity is independent of its X-ray morphology. In other words, it is possible that either merging cluster has not generated enough turbulence to form a radio halos or turbulence is already dissipated over a time scale $\sim$ Gyr after the merging process and hence absence of radio halo. In both of these conditions, host cluster's X-ray morphology would be disturbed.

\par There is a steepening in the spectrum of the MACSJ0417.5-1154 radio halo at $\sim$ 610 MHz. {It is possible that the halo is characterized by a rest-frame cut-off frequency at $\sim$ 900 MHz (610 MHz $\times$ (1+$z$))}. This cut-off frequency depends on the magnetic fields and acceleration efficiency in the ICM, and on the energy density of the cosmic microwave background (CMB) radiation \citep{2010A&A...517A..10C}. Further, it is believed that $\sim$ Gyr after cluster merger turbulence in the ICM becomes weaker and there are a strong synchrotron energy losses which produce a cut-off between $\sim$ 100 MHz and GHz frequency range \citep{2010A&A...517A..10C, 2011MmSAI..82..499V}. However, most of the massive clusters which host halos have cut-off frequencies $\geq$ 1.4 GHz, while in the case of MACSJ0417.5-1154 ($M_{500}$ = 22.1 $\times$ 10$^{14}$ $M\odot$) the cut-off is at $<$ GHz frequency. One possible scenario is that sufficient amount of time has passed since merger ($\sim$ 1 Gyr) and turbulence has decayed in the MACSJ0417.5-1154. In such a  scenario due to less efficient re-acceleration the cut-off frequency shifts to $<$ GHz frequencies. The position of the MACSJ0417.5-1154 in morphology parameter planes where it is situated at the edge of NR clusters which indicates less energetic merging events also supports this scenario. 
In earlier studies such ultra steep spectrum radio halos have been detected in A521 \citep{2008Natur.455..944B}, A1914 \citep{2003A&A...400..465B} and A2255 clusters \citep{1997A&A...317..432F}.   

\par
\begin{figure}
    \centering
    \caption {\small Plot showing the $L_{x}$ vs. $P_{1.4GHz}$ correlation for radio halos. The data from the current study are labeled with the respective cluster names. The remaining data are from the literature \citep{2015A&A...579A..92K}. 
The red filled circles  are confirmed halos while the blue squares are candidate halos. The arrows indicate upper limits. The solid line represents the best-fit for the $L_{x}$--$P_{1.4GHz}$ relation for halos (log($P_{1.4GHz}$) = A $\times$ log($L_{X}$) + B, where A = 2.24 $\pm$ 0.28 and B = -76.41 $\pm$ 12.65).}
   \label{P14lxplot}
   \includegraphics[origin=c,width=\columnwidth]{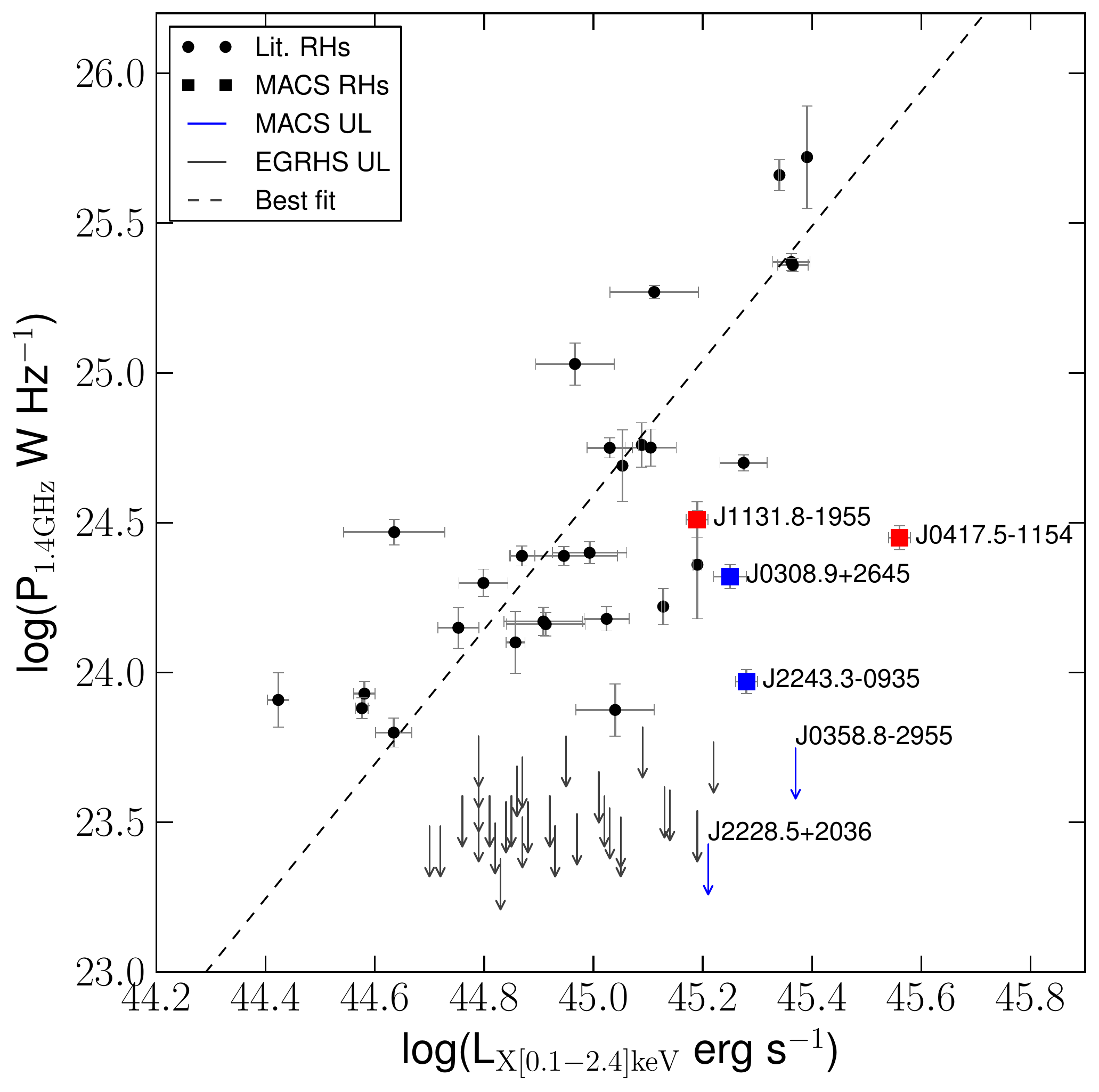}
\end{figure}

 We plotted radio powers at 1.4 GHz ($P_{1.4GHz}$) and the corresponding X-ray  luminosities (0.1--2.4 keV) of known halos in Figure~\ref{P14lxplot}. Since our sample clusters are at high redshifts we applied k-correction to their radio powers ($P_{1.4GHz}$). {For applying k-correction, we used the dimming factor of (1+$z$)$^{(1+\alpha)}$}. For calculating this, we used $\alpha$ = -1.72 for  MACSJ0417.5-1154, and $\alpha$ = -1.37 for MACSJ1131.8-1955, while for the remaining sample clusters we used an average spectral index of -1.3. Clusters  MACSJ1131.8-1955 (A1300) and MACSJ0417.5-1154, which host confirmed halos, and MACSJ0308.9+2645 and MACSJ2243.3-0935, which host candidate halos are situated close to the best-fitting line within intrinsic scatter of the $L_{X} - P_{1.4GHz}$ correlation. MACSJ0417.5-1154 is showing cut-off in the spectrum at the rest-frame frequency $\sim$ 900 MHz so obvious $P_{1.4GHz}$ is lower than its expected value. From the best-fit line it is lower by a factor of $\sim$ 11. If we use flux density value from the NVSS observation, then still it is lower by a factor of $\sim$ 10. The upper limit on the radio power of the MACSJ0358.8-2955 and MACSJ2228.5+2036 halos are lower by a factor of $\sim$ 11 compared to that expected from the best-fitting line. There could be an alternative to the primary model that secondary CR electrons occur from CR protons collisions with ambient thermal protons in a cluster volume. According to \cite{2011MNRAS.410..127B}, the relativistic electrons originating from hadronic ($p$-$p$) collisions could produce radio emission which is a factor $\sim$ 10 below the recent $P_{1.4GHz}$ vs. $L_{x}$ trend. Further, recent upper limits on radio luminosity of OFF state (non-detection) clusters based on GMRT observations constrain the ratio of CR protons and thermal energy density which is below several percent and magnetic field $\sim$ few $\mu$G \citep{2007ApJ...670L...5B, 2011ApJ...740L..28B}. These two latter clusters in our sample are also reaching to levels at which expected radio emission arise from secondary electrons produced in relativistic $p$-$p$ collisions. 

\par Future low frequency radio observations are crucial to shed new light on our current understanding of radio halos. In coming years, LOFAR, SKA1-LOW, MWA will increase the possibilities to utilise the frequency window at the bottom end of the radio spectrum, i.e. below $\sim$ 200 MHz. LOFAR will survey the sky at low frequencies with unprecedented depth and sensitivity to detect the faint and ultra-steep diffuse radio sources. Future SKA1 survey will provide us arcsec resolution with excellent surface brightness sensitivity at low frequencies. These future radio observations will permit us to study the formation and evolution of radio halos in an entirely new range of redshifts and cluster masses. Further improved sensitivities of these newer observations will have implications on the hardronic model.       

\section*{Acknowledgements} 
We thank anonymous referee for his/her critical comments that help us to improved this paper. RK is supported through the INSPIRE Faculty Award of the Department of Science and Technology, Government of India. We thank the staff of the GMRT who have made these observations possible. GMRT is run by the National Centre for Radio Astrophysics of the Tata Institute of Fundamental Research. The National Radio Astronomy Observatory is a facility of the National Science Foundation operated under cooperative agreement by Associated Universities, Inc. The scientific results reported in this article are based in part on data obtained from the Chandra Data Archive and published  previously in cited articles.

\bibliography{references}

\end{document}